\documentclass[iop]{emulateapj}

\usepackage{enumitem} 

\newcommand{\hMsun}{{\ifmmode{h^{-1}{\rm
        {M_{\odot}}}}\else{$h^{-1}{\rm{M_{\odot}}}$~}\fi}} 
\newcommand{\hMpc}{{\ifmmode{h^{-1}{\rm Mpc}}\else{$h^{-1}$Mpc }\fi}}
\def\be{\begin{equation}}
\def\ee{\end{equation}}
\def\ba{\begin{eqnarray}}
\def\ea{\end{eqnarray}}

\shorttitle{Reshift Dependent Alcock-Paczynski effect from SDSS-III DR12}
\shortauthors{X.-D. Li, C. Park, C.G. Sabiu, H. Park, D.H. Weinberg, J. Kim, S.E. Hong}

\begin{document}


\title{Cosmological constraints from the redshift dependence of the Alcock-Paczynski effect: application to the SDSS-III BOSS DR12 galaxies}



\author{Xiao-Dong Li, Changbom Park,}
\affil{School of Physics, Korea Institute for Advanced Study, 85 Heogiro, Dongdaemun-gu, Seoul 02455, Korea}
\author{Cristiano G. Sabiu, Hyunbae Park,}
\affil{Korea Astronomy and Space Science Institute, Daejeon 305-348, Korea}
\author{David H. Weinberg,}
\affil{Department of Astronomy and CCAPP, The Ohio State University, 140 West 18th Avenue, Columbus, OH 43210, USA}
\author{Donald P. Schneider,}
\affil{Department of Astronomy and Astrophysics, The Pennsylvania State University, 
University Park, PA 16802 }
\affil{Institute for Gravitation and the Cosmos, The Pennsylvania State University, 
University Park, PA 16802 }
\author{Juhan Kim\altaffilmark{1},}
\affil{Center for Advanced Computation, Korea Institute for Advanced Study, 85 Hoegi-ro, Dongdaemun-gu, Seoul 02455, Korea}
\affil{School of Physics, Korea Institute for Advanced Study, 85 Heogiro, Dongdaemun-gu, Seoul 02455, Korea}
\and
\author{Sungwook E. Hong}
\affil{School of Physics, Korea Institute for Advanced Study, 85 Heogiro, Dongdaemun-gu, Seoul 02455, Korea}

\altaffiltext{1}{Corresponding Author: kjhan@kias.re.kr}


%
%
%
%
%


\begin{abstract}
We apply the methodology developed in \cite{Li2014,Li2015} to BOSS DR12 galaxies and 
derive cosmological constraints from the redshift dependence of the Alcock-Paczynski (AP) effect.
The apparent anisotropy in the distribution of observed galaxies arise from two main sources,
the redshift-space distortion (RSD) effect due to the galaxy peculiar velocities,
and the geometric distortion when incorrect cosmological models are assumed for 
transforming redshift to comoving distance,
known as the AP effect.
Anisotropies produced by the RSD effect are, although large,
maintaining a nearly uniform magnitude over a large range of redshift,
while the degree of anisotropies from the AP effect varies with redshift by much larger magnitude. 
We split the DR12 galaxies into six redshift bins, measure the 2-point correlation function in each bin,
and assess the redshift evolution of anisotropies.
We obtain constraints of 
$\Omega_m=0.290 \pm 0.053,\ \ w = -1.07 \pm 0.15$,
which are comparable with the current constraints from other cosmological probes
such as type Ia supernovae, cosmic microwave background, and baryon acoustic oscillation (BAO).
Combining these cosmological probes with our method yield tight constraints of 
$ \Omega_m = 0.301 \pm 0.006,\ w=-1.054 \pm 0.025$.
Our method is complementary to the other large scale structure probes like BAO and topology.
We expect this technique will play an important role in deriving cosmological constraints from
large scale structure surveys.
\end{abstract}





\section{Introduction}

The current standard model of cosmology has been highly successful at reproducing the Universe on large scales. 
From the temperature fluctuations in the cosmic microwave background (CMB), 
to the late time clustering of galaxies, 
the vacuum energy dominated cold dark matter model ($\Lambda$CDM) fits the data surprisingly well \citep{Planck2015,Anderson2013}. 
This result is all the more impressive considering both the underlying assumptions, 
such as homogeneity, isotropy, scale invariance of the primordial fluctuations, 
and the minimal set of cosmological parameters that are required.

Nonetheless, these models produce the unsatisfactory prospect that we must include within our ontology both a vacuum energy 
that is much smaller than that predicted from  quantum mechanics, 
or alternatively a new scalar field (dark energy) that has negative pressure
\citep{SW1989,Riess1998,Perl1999,PR2003,Li2011}, 
and a new matter component, which is not contained within the standard $SU(3)\times SU(2) \times U(1)$ formulation of particle physics.

With an over-abundance of models for both dark energy-like accelerated expansion and dark matter, 
it is crucial to obtain precise and model-independent measurements of the cosmic evolution, usually referred to as background observables. 
Two such observables are the angular diameter distance, $D_A$, and the Hubble factor, $H$.  
If these quantities can be measured at various redshifts and to a high degree of accuracy then our ability to differentiate between various competing models will be greatly increased.

In the last few years there has been increasing interest in using the Alcock-Paczynski (AP) effect \citep{AP1979} 
in the large-scale clustering of galaxies to obtain constraints on $D_A$ and $H$ \citep{Guzzo2008,topology}. 
Assuming an incorrect cosmological model for the coordinate transformation from redshift space to comoving space produces residual geometric distortions. 
These distortions are induced by the fact that measured distances along and perpendicular to the line of sight are fundamentally different. 
Measuring the ratio of galaxy clustering in the radial and transverse directions provides a probe of this AP effect.

There have been several methods proposed for applying the AP test to the large scale structure (LSS).
The most widely adopted one uses anisotropic clustering \citep{Ballinger1996,Matsubara1996},
which has been used for the 2 degree Field Quasar Survey \citep{Outram2004},
the WiggleZ dark energy survey \citep{Blake2011}, 
the Sloan Digital Sky Survey-I/II (SDSS-I/II) Luminous Red Galaxy (LRG) survey \citep{Eisenstein et al. 2011,ChuangWang2012},
and the SDSS-III Baryon Oscillation Spectroscopic Survey (BOSS) 
\citep{Reid2012,Beutler2013,Linder2013,2014arXiv1407.2257S, 2014ApJ...781...96L,
Alam2016, Beutler2016, Sanchez2016}
The main caveat of this method is that,
because the radial distances of galaxies are inferred from redshifts,
AP tests are inevitably limited by redshift-space distortions (RSD) \citep{Ballinger1996},
which leads to apparent anisotropy even if the adopted cosmology is correct.
The RSDs must be accurately modeled for the 2-point statistics of galaxy clustering.

\cite{Marinoni2010} proposed using the symmetry properties of galaxy pairs.
Unfortunately this method is also seriously limited by RSD.
The peculiar velocity distorts the redshift and changes the apparent tilt angles of galaxy pairs.
The effect depends on both redshift and underlying cosmology, and is rather difficult to model accurately \citep{Jennings2011}.
\cite{Ryden1995} and \cite{LavausWandelt1995} proposed another method using the apparent stretching of voids.
This approach has the advantage that the void regions are easier to model compared with dense regions,
but has limitations in that it utilizes only low density regions of the LSS
and requires large samples.

\cite{Li2014} proposed another method utilizing the redshift dependence of AP effect to overcome the RSD problem.
The anisotropies produced by RSD effect are, although very large,
close to uniform in magnitude over a large range of redshift.
Conversely, if cosmological parameters are incorrectly chosen, the LSS
appear anisotropic and the degree of anisotropy varies with redshift. 
We used the {\it galaxy density gradient field} 
to characterize the anisotropies in LSS and tested the idea on Horizon Run 3 (HR3) N-body simulations \citep{horizonrun},
demonstrating that the method leads to unbiased estimation of the density parameter $\Omega_m$ and the dark energy equation of state (EoS) $w$.

The same topic was revisited in \cite{Li2015}, but using the {\it galaxy two-point correlation function} (2pCF) as the statistical tool.
The 2pCF as a function of angle, $\xi(\mu)$, is measured at different redshifts.
Similar to \cite{Li2014}, we found that the RSD effect, although significantly distorting $\xi(\mu)$, 
exhibits much less redshift evolution compared to the amount of change in $\xi(\mu)$ due to incorrectly adopted cosmologies.
When incorrect cosmological parameters are adopted, 
the shape of $\xi(\mu)$ appears anisotropic due to the AP effect, 
and the amplitude is shifted by the change in comoving volume;
both effects have significant redshift dependence.
We test the method using the 2pCF on mock surveys drawn from HR3
and find the constraints obtained are tighter than those from the methodology of \cite{Li2014}.

The change of the comoving volume size is another consequence of an incorrectly adopted cosmology, 
and has motivated investigations constraining cosmological parameters from number counting of 
galaxy clusters \citep{PS1974,VL1996}. 
An obstacle in using the comoving volume for cosmological tests is the evolution of the number of target objects.
The essential need for reducing the evolution effects in applying the test led \cite{topology} to propose a new method
using the topology of LSS. 
Since the topology is a measure of intrinsic connectivity of structures,
it is expected to be insensitive to non-linear gravitational evolution, type of density tracers,
and RSD on large scales.
This method has been applied to the WiggleZ Dark Energy Survey data by \cite{WiggleZtopoloy},
and to simulated BOSS samples \citep{Speare2015}.

In this paper we apply our methodology to SDSS-III BOSS Data Release 12 (DR12) galaxies \citep{Reidetal:2016}.
We take the 2pCF as statistical tool characterizing the anisotropic clustering and follow the procedure of \cite{Li2015} to conduct the analysis.
We assume a flat Universe and constrain parameters of $\Omega_m$ and $w$.
2pCF is a mature statistic in cosmology and its optimal estimation and statistical properties are well understood.
Compared with the density gradient field statistic, it leads to tighter constraints and is less affected by survey geometry.

The outline of this paper is as follows. 
In Sec. 2 we describe the observational data used in this paper.
In Sec. 3 we discuss the N-body simulations and mock galaxy catalogues that are used in this analysis.
In Sec. 4 we briefly review the nature and consequences of the AP effect when performing coordinate transforms in a cosmological context. 
In Sec. 5 and Sec. 6, we describe our analysis method and present the cosmological constraints obtained from BOSS DR12 galaxies. 
We conclude in Sec. 7.

\section{The Observational Data}\label{sec:data}

The Sloan Digital Sky Survey (SDSS; York et al. 2000) 
imaged approximately 7\,606 $\rm deg^2$
of the Northern Galactic Hemisphere and 
3\,172 $\rm deg^2$ of the Southern Galactic Hemisphere in the {\it ugriz} bands \citep{Fukugita1996,Gunn1998}.
The survey was performed using the 
2.5m Sloan telescope \citep{Gunn et al. 2006}
at the Apache Point Observatory in New Mexico.
BOSS \citep{Dawson et al. 2012,Smee2013}, as a part of the SDSS-III survey \citep{Eisenstein et al. 2011},
has obtained spectra and redshifts 
of 1.37 million galaxies selected from the SDSS imaging,
covering a region of 9\,376 $\rm deg^2$.
The galaxy redshifts were measured by an automated pipeline \citep{Bolton2012}.

\begin{figure*}
   \centering{
   \includegraphics[width=8.5cm]{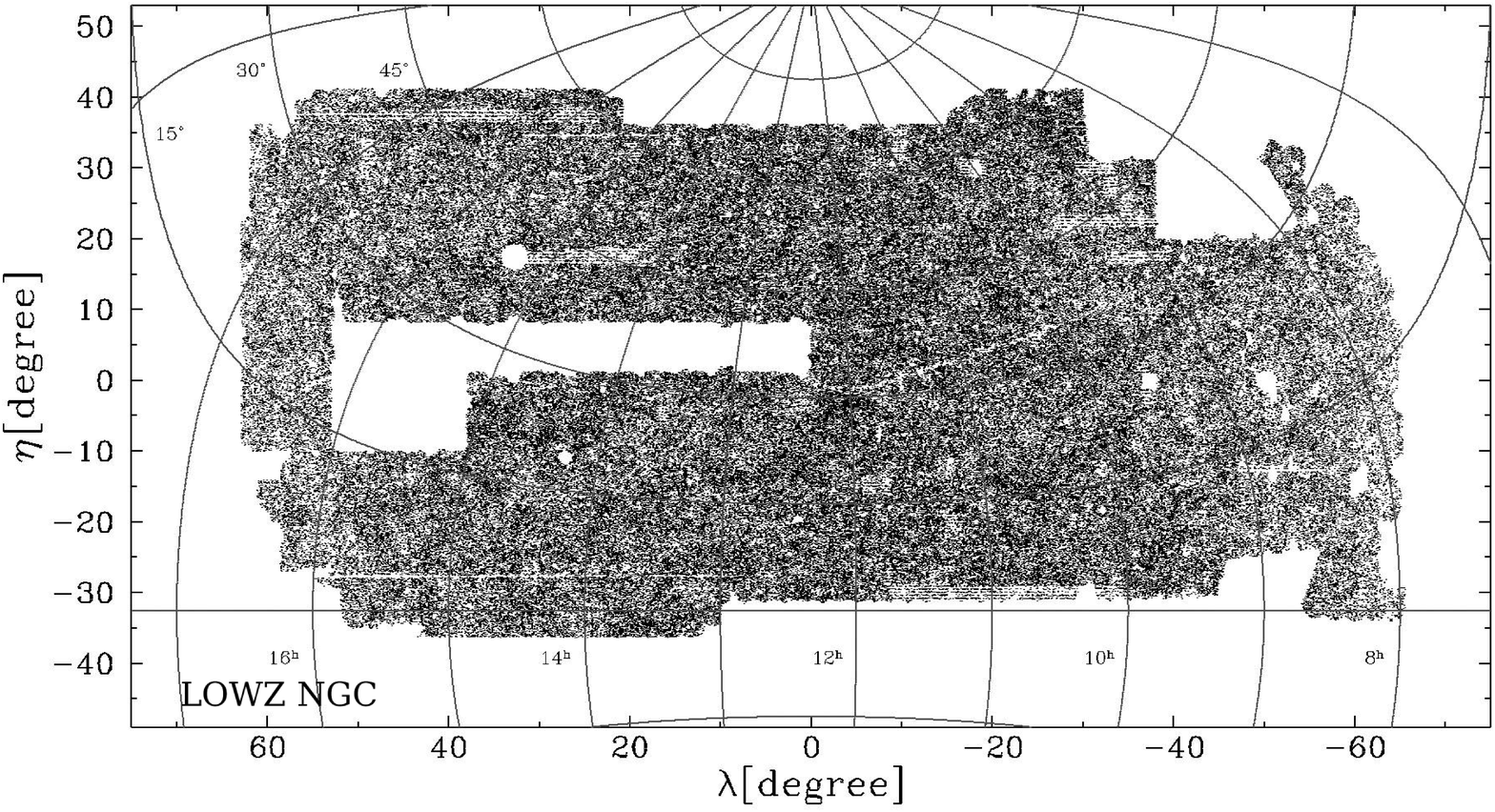}
   \includegraphics[width=8.5cm]{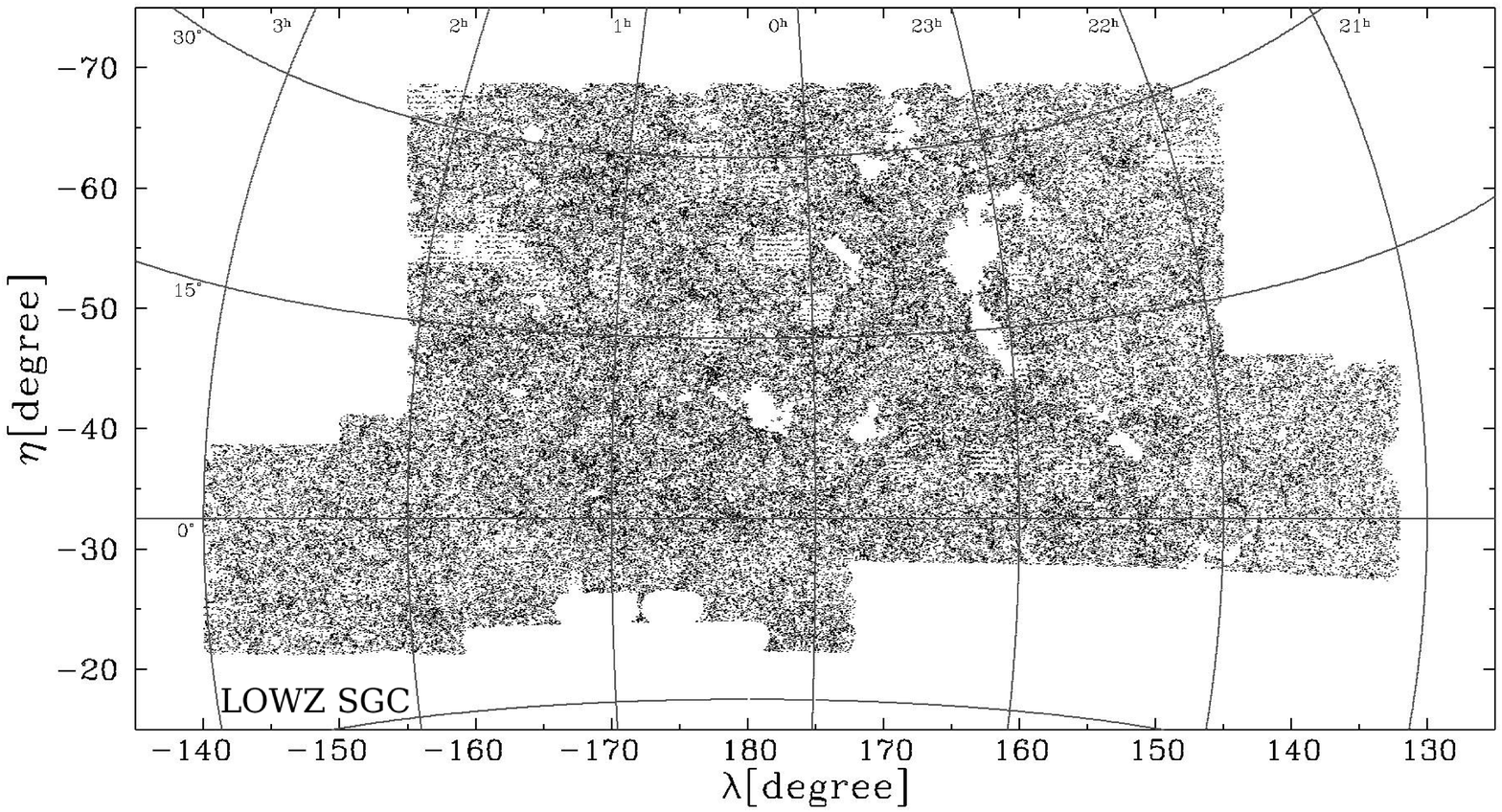}
   \includegraphics[width=8.5cm]{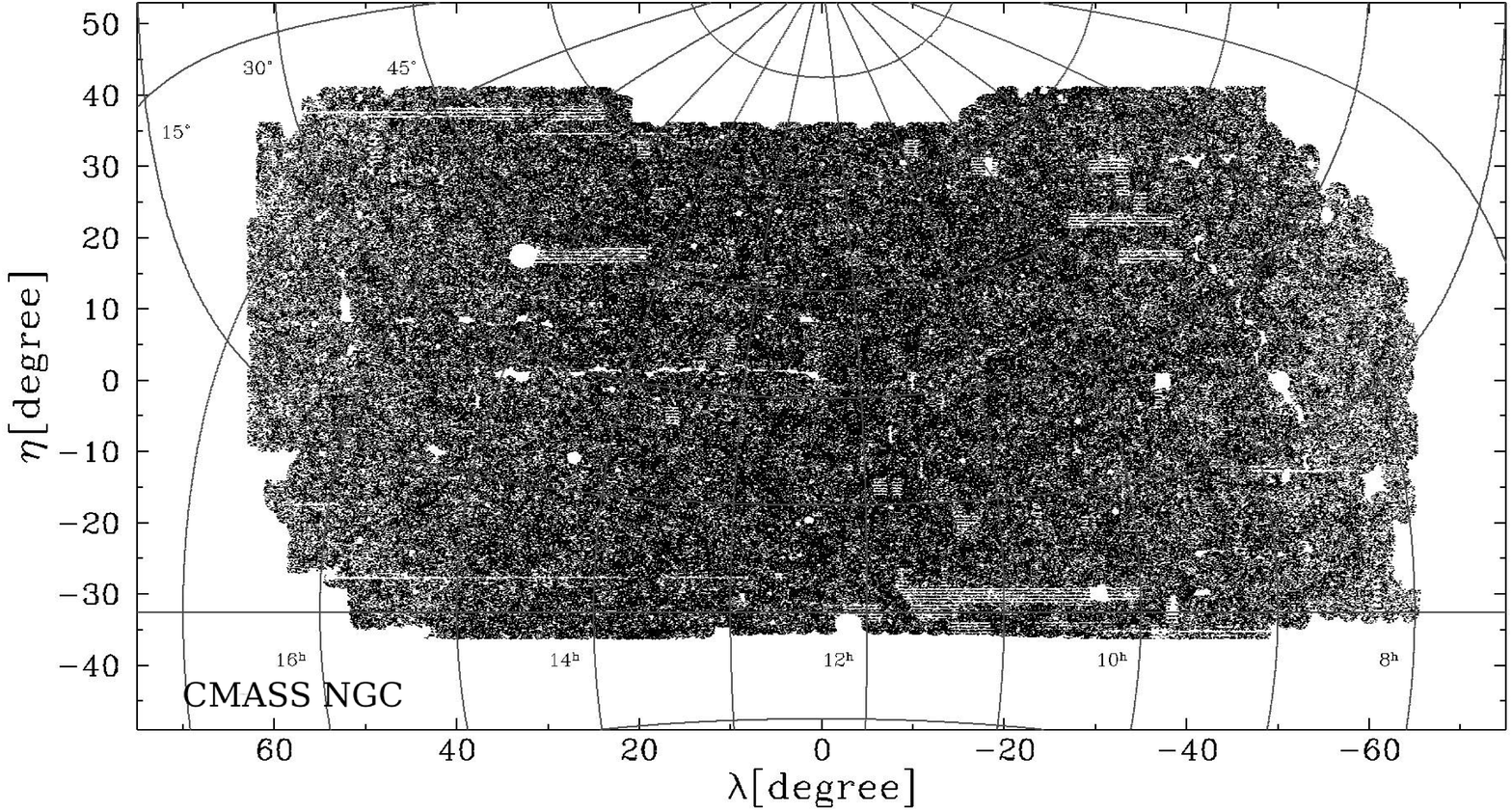}
   \includegraphics[width=8.5cm]{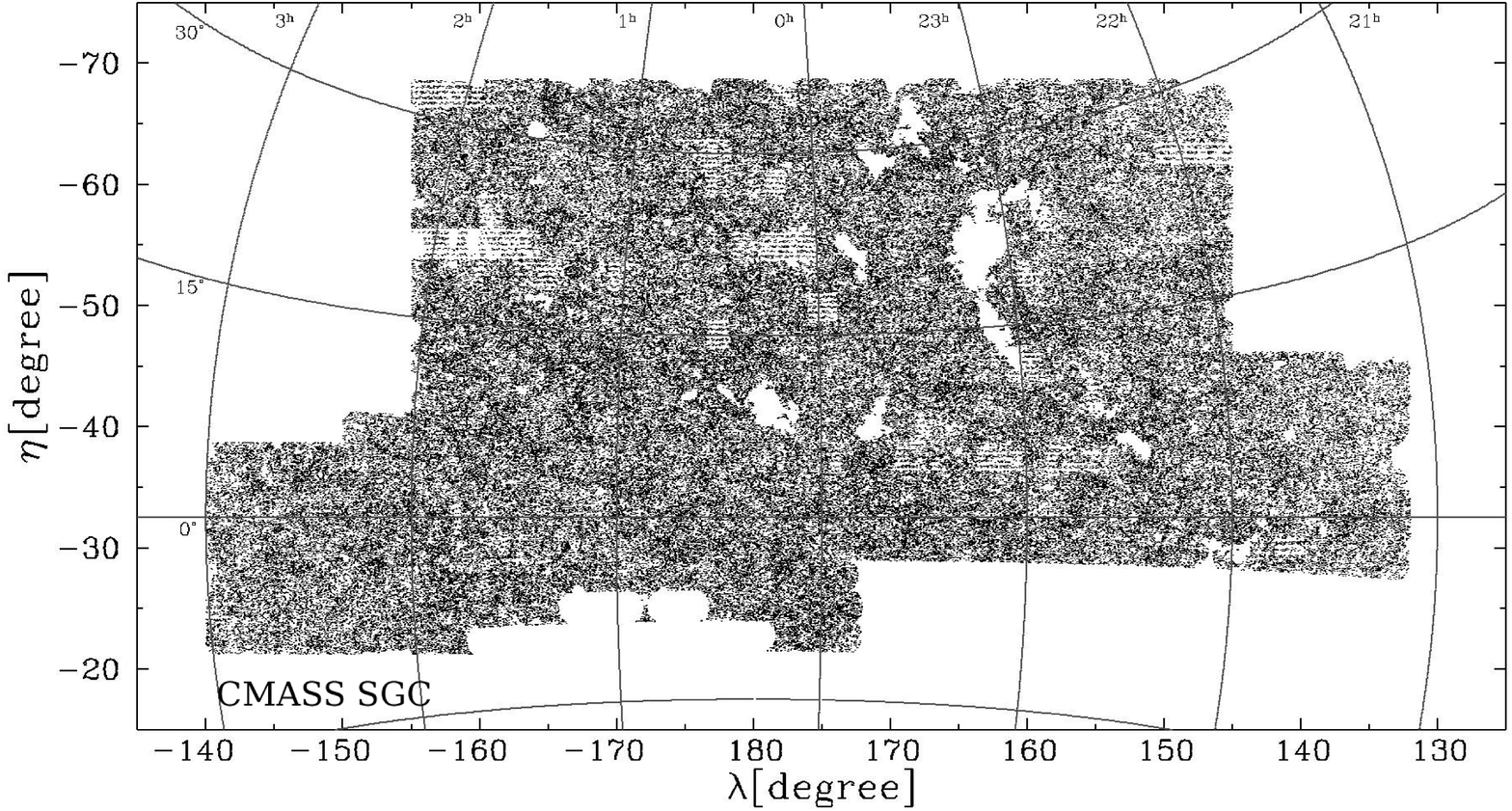}
   }
   \caption{\label{fig_radec}
   The sky coverage of the LOWZ and CMASS samples in the north and south Galactic caps. 
   The individual points mark position of galaxies in the survey coordinate frame.
   The solid lines mark the right ascension and declination.
   The mean completeness is 97.2\% for the LOWZ sample, shown in the upper panels,
   and 98.8\% for the CMASS sample in the lower panels.
   The effective sky coverage is 8,337 $\rm deg^2$ for LOWZ and 9,376 $\rm deg^2$ for CMASS.
   See \cite{Reidetal:2016} for more details.
   }
\end{figure*}

The spectroscopic sample of BOSS has two primary catalogues.
The LOWZ sample is designed to extend the SDSS-I/II LRG sample to $z\approx 0.4$ and fainter luminosities,
in order to increase the number density of the sample by a factor of 3.
The CMASS sample covers a higher redshift ($0.4\lesssim z \lesssim 0.7$).
It was targeted to be an approximately stellar mass limited sample of massive, luminous galaxies.
The final data release (DR12) samples are described in \cite{Reidetal:2016},
where the details of targeting algorithms and the catalogues are provided.

\begin{figure*}
   \centering{
   \includegraphics[width=19cm]{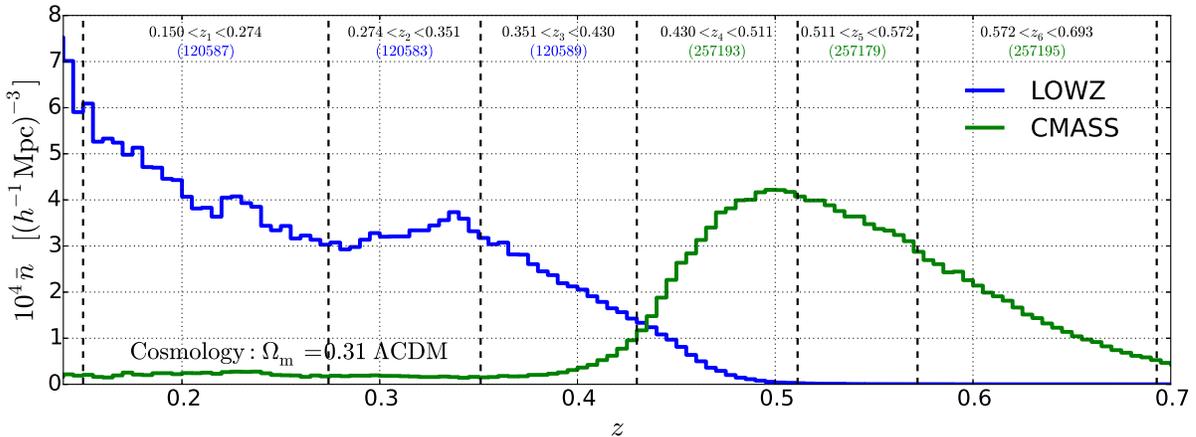}
   }
   \caption{\label{fig_nbar}
      The redshift density distribution of the BOSS DR12 galaxy samples, assuming a $\Lambda$CDM cosmology with $\Omega_m=0.31$.
      The blue and green solid histograms show the distribution of LOWZ and CMASS galaxies respectively. 
      The vertical dashed lines define the six redshift bins that are used to cut the samples.
      Their redshift ranges are listed.
      The number of LOWZ (CMASS) galaxies in the three low (high) redshift bins are presented in the brackets.
      }
\end{figure*}

Figure \ref{fig_radec} presents the sky coverage of the BOSS DR12 samples used in this analysis. 
The mean completeness is 97.2\% for the LOWZ sample, in the upper panels, and 98.8\% for the CMASS sample shown in the lower panels.
Figure \ref{fig_nbar} shows the galaxy number density of the two samples.
In this analysis we use 361\,759 LOWZ galaxies at $0.15<z <0.43$ and 
771\,567 CMASS galaxies at $0.43< z < 0.693$.
We split the galaxies into six redshift bins: 
three bins in LOWZ ($0.150<z_1<0.274<z_2<0.351<z_3<0.430$), 
and three in CMASS ($0.430<z_4<0.511<z_5<0.572<z_6<0.693$), 
as illustrated in Figure \ref{fig_nbar}.

\begin{figure*}
   \centering{
   \includegraphics[width=19cm]{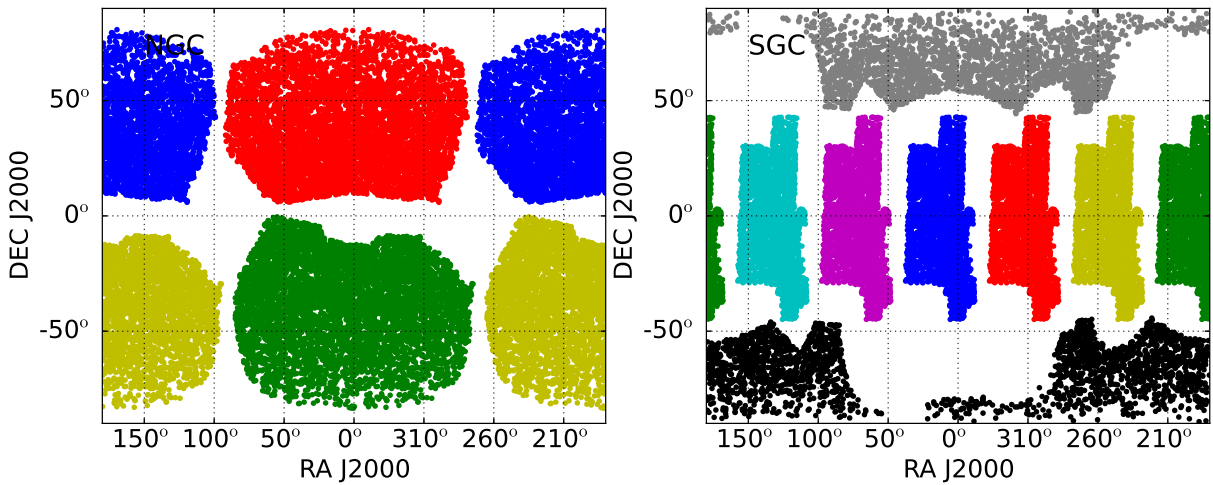}
   }
   \caption{\label{fig_mock}
      Creation of galaxy samples for BOSS, from the HR3 or HR4 simulations.
      From an all-sky mock survey,
       we are able to produce four sets of NGC samples or eight sets of SGC samples with non-overlapping sky coverage.
      The individual points are the right ascension and declination of 1\% galaxies drawn from the CMASS samples.
      }
\end{figure*}

Each spectroscopically observed galaxy is assigned several weights to account for observational effects. 
The galaxy weights are constructed from three distinct effects.
Firstly, galaxies lacking a redshift due to fiber collisions 
\footnote{Fiber collisions occur when two objects are close enough together such that two fibers cannot be placed.
In BOSS, the collision radius is 62 $''$.}
or inadequate spectral information are accounted for by reweighting
the nearest galaxy by a weight $w_{\rm fail}=(1+n)$, 
where $n$ is the number of close neighbors without a measured redshift. 
Secondly, all galaxies are assigned `FKP' weights \citep{1994ApJ...426...23F}
as a function of number density,
to optimize the clustering measurements in the face of shot-noise and cosmic variance.
The third weight corrects for angular variations of survey completeness and the systematics 
related to the angular variations in stellar density that make detection of
galaxies difficult in over-crowded regions of the sky. 
The total weight for each galaxy is the product of these three weights, $w_{\rm total}=w_{\rm fail}w_{\rm FKP}w_{\rm sys}$. 

For the statistical analyses, 
random catalogues having the same angular and redshift selection functions as the data 
are provided along with the data \citep{Reidetal:2016}.
The random points are also weighted but they only include the minimum variance `FKP' weight. 

\section{The Mock Galaxy Data}
\label{sec:mocks}


For LSS studies mock survey samples created from simulations are crucial
for the correction of systematics and covariance estimation.
The Horizon Run simulations are a suite of large volume N-body simulations that 
have resolutions and volumes capable of accurately reproducing the observational statistics of the current major redshift surveys like 
SDSS-III BOSS \citep{park 2005,2009ApJ...701.1547K,horizonrun}.
The HR3 \citep{horizonrun} and HR4 \citep{hr4} simulations, and the MultiDark-Patchy mock catalogues \citep{MDPATCHY} are used in our analysis.

From the HR3 and HR4 simulations we have generated all-sky light cone mock galaxy catalogues.
The all-sky spherical mocks are then incorporated with the same fiber collision effect, 
angular masks, and radial selection function with the real observational data,
creating mock surveys of BOSS DR12.

We impose a minimum mass limit varying along with redshift to match the radial density of BOSS samples.
The galaxies of the BOSS DR12 samples do not cleanly distribute above some particular mass boundary;
they always have a fuzzy, blur boundary extending to relatively small values (as an example, see Figure 3 of 
Parihar et al. 2014). 
Galaxies in the mock samples are systematically more massive than those from observations;
We will discuss the effect of this discrepancy in Sec. \ref{sec:caveats}.

\subsection{Horizon Run 4}

The HR4 simulation  \citep{hr4} used a box size $L={3150}$ $h^{-1}$Mpc, 
and $N=6300^3$ particles.  
The simulation used the second order Lagrangian perturbation theory (2LPT) initial conditions at $z_{i}=100$ and a WMAP5 cosmology
$(\Omega_{b},\Omega_{m},\Omega_\Lambda,h,\sigma_8,n_s)$  = (0.044, 0.26, 0.74, 0.72, 0.79, 0.96) \citep[]{komatsu 2011}, 
yielding a particle mass of $m_{p} \simeq 9.02 \times 10^9 \hMsun$.
This starting redshift, combined  with 2LPT initial conditions, ensures an accurate mass function and power spectrum \citep{2014NewA...30...79L}. 

Mock galaxy samples are produced from the HR4 simulation by using a modified version of the one-to-one correspondence scheme \citep{hong2016}. 
The most bound member particles (MBPs) of simulated halos are adopted as the tracer of galaxies rather than the subhalos,
and the merger timescale is taken into account in the lifetime of merged halos.
We built the merger trees of simulated halos by tracking their MBPs from $z = 12$ to 0.
When a merger event occurs, we calculate the merger timescale described in \cite{jiang2008} (hereafter J08)
to determine when a satellite galaxy is completely disrupted.
By using the abundance matching, 
we modeled the luminosity of a central/isolated galaxies from their current mass
and of satellite galaxies 
from their mass at the time of infall.

\cite{hong2016} compared the 2pCF of our mock galaxy sample at $z = 0$ to the SDSS DR7 volume-limited galaxy sample \citep{zehavi2011}.
The simulated 2pCF shows a similar finger of god (FOG) feature \citep{FOG} as the observation in the contour map, 
and the projected 2pCF agrees with the observation within 1$\sigma$ deviation on scales greater than 1 ${h^{-1}}$Mpc.

The HR4 simulation yield one all-sky light cone mock galaxy catalogue reaching $r=3\,150$ ${h^{-1}}$Mpc.
As shown in Figure \ref{fig_mock},
from one all-sky survey, 
we are able to create four sets of non-overlapping north Galactic cap (NGC) samples for CMASS and LOWZ, 
or eight sets of non-overlapping south Galactic cap (SGC) samples.
In this paper we use these simulated galaxies for the estimation of the systematics in the 2pCF of the observed galaxies.

\begin{figure*}
   \centering{
   \includegraphics[height=8cm]{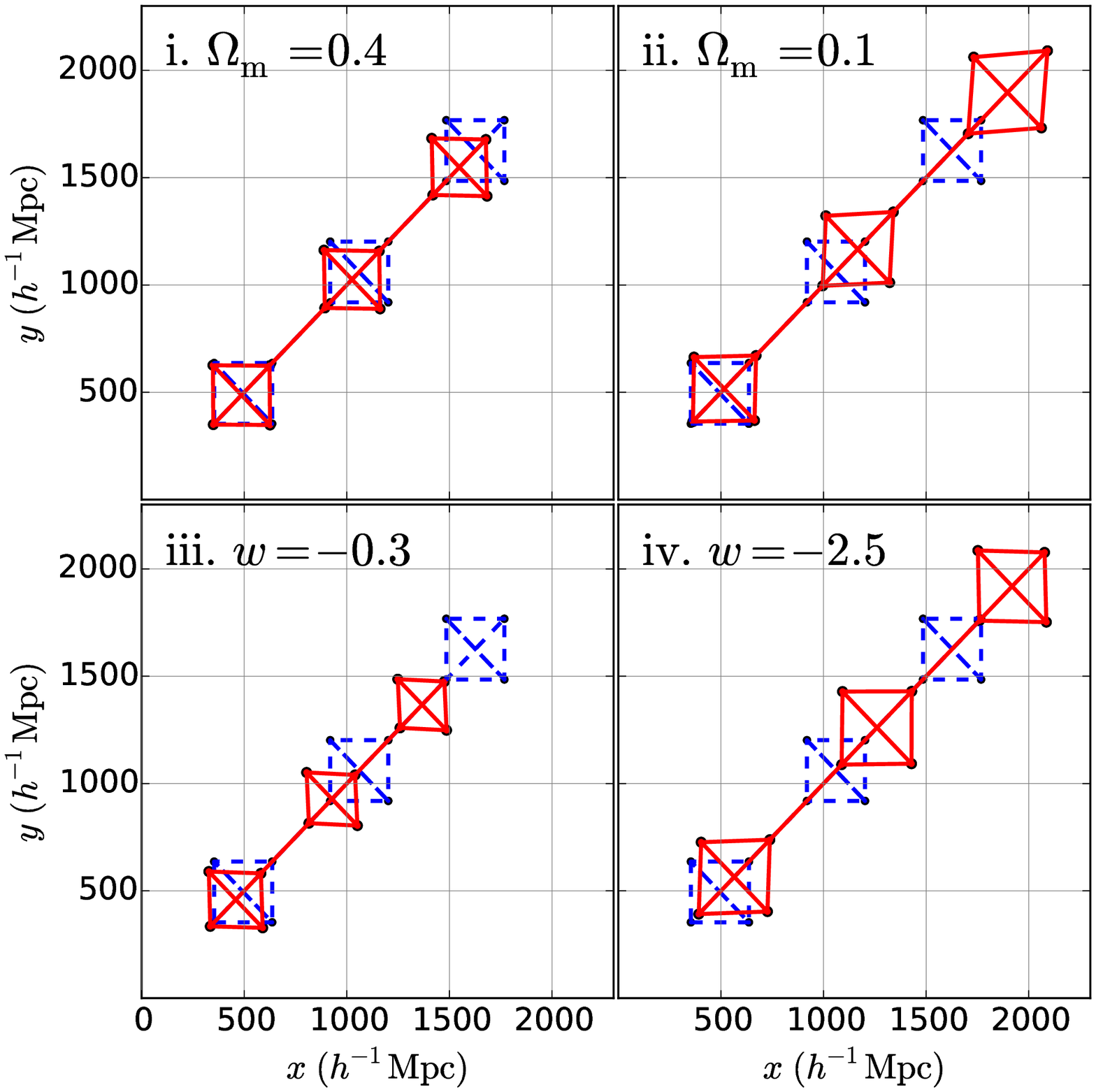}
   \includegraphics[height=8cm]{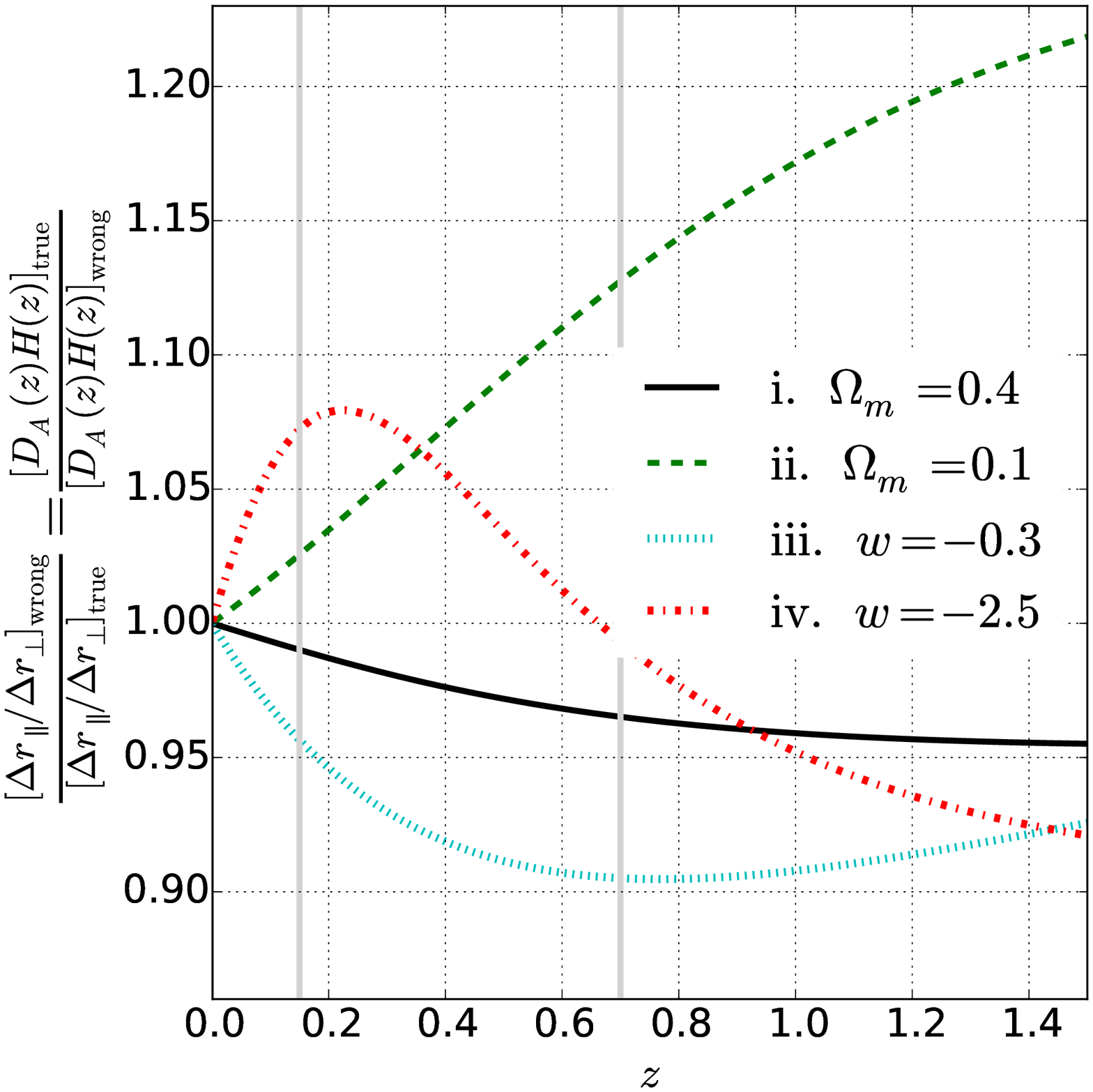}
   }
   \caption{\label{fig_xy}
   The redshift dependence of the AP effect in four incorrect cosmologies,
   assuming that the true cosmology is $\Omega_m=0.31$, $w=-1$.
   The left panel shows the apparent distortion of four perfect squares,
   measured by an observer located at the origin.
   The apparently distorted shapes are plotted in red solid lines.
   The underlying true shapes are indicated in blue dashed lines.
   The right panel displays the degree of the shape distortion, as described by Equations (\ref{eq:stretch}).
   The BOSS DR12 galaxies used in our analysis have a redshift coverage of $0.15 < z < 0.693$ (marked 
   by the gray vertical lines). Clearly, the magnitude of the shape distortion due to AP changes with redshift.
   }
\end{figure*}

\subsection{Horizon Run 3}

As in HR4, HR3 also adopts a flat-space $\Lambda$CDM cosmology with the WMAP 5 year parameters.
The simulation was made within a cube of volume $(10.815\  h^{-1}{\rm{Gpc}})^3$
using $7120^3$ particles with particle mass of $1.25\times 10^{11}$\hMsun.
The simulations were integrated from $z=27$ and reached $z=0$ after making $N_{\rm step}=600$ global timesteps.
The collapsed high-density regions were identified using the Friend-of-Friend algorithm with the linking length of 0.2 times the mean particle separation.
The physically self-bound (PSB) subhalos that are gravitationally self-bound and tidally stable \citep{kim and park 2006} 
are identified and used as galaxy proxies.
The PSB halo finder is a group-finding algorithm that can efficiently identify halos located even in crowded regions. 
This method combines two physical criteria such as the tidal radius of a halo and the total energy of each particle to find member particles.
The group velocity of member particles is adopted as the peculiar velocity of each PSB subhalo.



We generate multiple BOSS-like surveys by placing 27 evenly spaced
observers within the HR3 cubical volume and allowing each to 
survey out to a redshift of 0.7
\footnote{In the analysis we choose the maximal redshift as 0.693 rather than 0.7. 
The outer boundary of the mock survey becomes fuzzy due to the peculiar velocity effect 
on the galaxy redshifts (Eq. (\ref{eq:zvpeu})).
A population of galaxies, that are expected to enter the $z<0.7$ region from the outside, is missing.
To avoid this problem we set the maximal redshift at 0.693, 
23.3 $h^{-1}{\rm Mpc}$ away from $z=0.7$.}.
Each of these 27 independent and non-overlapping spherical 
regions are further cut up into the required SDSS survey geometry,
resulting in 72 non-overlapping light-cone galaxy catalogues 
\footnote{Using the 27 spherical light cones, we create 72 sets of NGC samples from 18 light cones, 
and 72 sets of SGC samples from the other 9 light cones.}
simulating BOSS DR12 within the redshift range of $0.15< z< 0.693$.
These mock surveys are used to estimate the covariance of the 2pCF in our analysis.

\subsection{MultiDark-Patchy Mocks}

In additional to the HR3 mock surveys, 
we also use 2\,000 MultiDark-Patchy mock catalogues \citep{Kitaura2014,MDPATCHY} to estimate the covariance matrix.

The MultiDark-Pathy mocks were produced using approximate gravity solvers 
and analytical-statistical biasing models.
They have been calibrated to a BigMultiDark N-body simulation \citep{K2014}, which 
uses $3\,840^3$ particles in a volume of $(2.5h^{-1}\rm Gpc)^3$
assuming a $\Lambda$CDM cosmology with
 $\Omega_m = 0.307115$, $\Omega_b = 0.048206$, $\sigma_8 = 0.8288$, $n_s = 0.9611$, and $H_0 = 67.77 {\rm km}\ s^{-1} {\rm Mpc}^{-1}$.
Halo abundance matching is used to reproduce the two and three-point 
clustering measurements of BOSS \citep{RT2015}.
The redshift evolution of the biased tracers 
is matched to observations by applying the 
aforementioned technique in a number of redshift bins,
with the resulting mock catalogues being combined together 
to form a contiguous lightcone.

The resulting MultiDark-Patchy mock surveys reproduce the number density, 
selection function, survey geometry of the BOSS DR12 catalogues.
The 2pCF of the observational data is reproduced down to a few Mpc scales, in general within $1\sigma$ \citep{MDPATCHY}.
The MultiDark-Patchy mocks have been adopted for 
statistical analysis of BOSS data
in a series of works (see \cite{Alam2016} and references therein).
This large set of mock catalogues enabled us to perform 
a robust error estimation of the 2pCFs measured from the BOSS DR12 galaxies.

\section{AP Effect in Incorrect Cosmologies}
\label{sec:APeffect}

This section illustrates the AP effect when an incorrect cosmological model is used to calculate the distances of galaxies. Similar illustrations have been provided in \cite{Li2014,Li2015}.

Suppose that we probe the shape and volume of an object in the Universe,
which spans $\Delta z$ in redshift and $\Delta \theta$ in angle.
Its comoving sizes in the radial and transverse directions are given by
\begin{eqnarray}\label{eq:distance}
& &\Delta r_{\parallel} = \frac{c}{H(z)}\Delta z,\nonumber\\
& &\Delta r_{\bot}=(1+z)D_A(z)\Delta \theta,
\end{eqnarray}
where $H$ is the Hubble parameter and $D_A$ the proper angular diameter distance.
In the particular case of a flat Universe with constant dark energy EoS, they take the forms of
\begin{eqnarray}\label{eq:HDA}
& &H(z) = H_0\sqrt{\Omega_ma^{-3}+(1-\Omega_m)a^{-3(1+w)}},\nonumber\\
& &D_A(z) = \frac{c}{1+z}r(z)=\frac{c}{1+z}\int_0^z \frac{dz^\prime}{H(z^\prime)},
\end{eqnarray}
where $a=1/(1+z)$ is the cosmic scale factor,
$H_0$ is the present value of Hubble parameter and $r(z)$ is the comoving distance.

In case we adopted an incorrect set of cosmological parameters in Equation (\ref{eq:HDA}),
the inferred $\Delta r_{\parallel}$ and $\Delta r_{\bot}$ are also incorrect,
resulting in distorted shape (AP effect) and wrongly estimated volume (volume effect).
The degree of variations in shape and volume are
\begin{equation}\label{eq:stretch}
 \frac{[\Delta r_{\parallel}/\Delta r_{\perp}]_{\rm wrong}}{[\Delta r_{\parallel}/\Delta r_{\perp}]_{\rm true}} =
  \frac{[D_A(z)H(z)]_{\rm true}}{[D_A(z)H(z)]_{\rm wrong}} 
\end{equation}
\begin{equation}\label{eq:volume}
 \frac{{\rm Volume}_{\rm wrong}}{{\rm Volume}_{\rm true}}
 = \frac{[\Delta r_{\parallel}(\Delta r_{\perp})^{2}]_{\rm wrong}}{[\Delta r_{\parallel}(\Delta r_{\perp})^{2}]_{\rm true}}
 = \frac{[D_A(z)^2/H(z)]_{\rm wrong}}{[D_A(z)^2 / H(z)]_{\rm true}},
\end{equation}
where ``true'' and ``wrong'' denote the values of quantities in the true cosmology and incorrectly assumed cosmology.
From the AP and volume effects, we can constrain  $D_A(z)H(z)$ and $D_A(z)^2 / H(z)$, respectively.

The apparent distortion of objects due to incorrect cosmological parameters is illustrated in the left panel of Figure \ref{fig_xy}.
Suppose that the true cosmology is a flat $\Lambda$CDM model with the present density parameter $\Omega_m=0.31$
and standard dark energy EoS $w=-1$ \citep[the best $\Lambda$CDM model determined by Planck 2015 results][]{Planck2015}.
If we distributed three square objects at various distances from 500 to 2,000 $h^{-1}$Mpc,
and had an observer located at the origin measure their redshifts and compute their positions and shapes 
using redshift-distance relations of four incorrect cosmologies:
\begin{enumerate}[label=(\roman*)]
\item $\Omega_m=0.4$, $w=-1$,
\item $\Omega_m=0.1$, $w=-1$,
\item $\Omega_m=0.31$, $w=-0.3$,
\item $\Omega_m=0.31$, $w=-2.5$,
\end{enumerate}
the mismatch between the true and assumed cosmology will cause the shapes of the squares appear distorted.
In the cosmological models (ii) and (iv) the squares are stretched in the line of sight (LOS) direction (hereafter ``LOS shape stretch''),
while in the models (i) and (iii) we see opposite effects of LOS shape compression.

The right panel of Figure \ref{fig_xy} presents the degree of shape distortion as a function of redshift. 
In cosmology (i) and (iii), $\frac{[D_A(z)H(z)]_{\rm true}}{[D_A(z)H(z)]_{\rm wrong}}$ have values less than 1, 
indicating LOS shape compression,
while in cosmology (ii) the curve lies above 1, 
corresponding to an LOS shape stretch.
The effect in cosmology (iv) is more subtle. 
There is a transition from LOS shape stretch to compression at $z \approx 0.65$.

More importantly, Figure \ref{fig_xy} highlights the redshift dependence of the AP effect. 
If the conversion of redshift to the comoving distance was correctly made,
there would be no shape distortion at any redshift.
Conversely, the four cases with incorrectly chosen cosmological parameters illustrated 
in Figure \ref{fig_xy} 
show characteristic dependence of the shape distortion on redshift.
We measure the 2pCF of BOSS DR12 galaxies in various redshift bins
and constrain cosmological parameters using the redshift evolution
of the anisotropic clustering signal.

\section{Methodology}\label{sec:methodology}

We measure the 2pCF in redshift bins of BOSS DR12 galaxies and 
determine cosmological parameters by examining the redshift evolution of clustering anisotropy.
Mock survey samples are used to correct the results for the systematics and to estimate the covariance.

\subsection{Grid of cosmology parameters}

The observed coordinates ({\it RA, Dec, z}) of galaxies
need to be converted to comoving coordinates ({\it x, y, z}) for the 2pCF analysis.
The dependence of clustering anisotropy on cosmology enters through the conversion from redshift to comoving distance,
i.e. the distance-redshift relation $r(z)$.
We consider the case of a flat Universe dominated by matter and dark energy,
so our $r(z)$ is governed by two parameters, $\Omega_m$ and $w$, as presented in Equation (\ref{eq:HDA})	.

In constraining these two parameters we examine the parameter space of 
$0.06\leq \Omega_m\leq 0.41$ and $-1.5 \leq w \leq -0.4$ with intervals of 
$\delta \Omega_m = 0.005$ and $\delta w = 0.025$,
forming a 71$\times$45 grid.
For each set of ($\Omega_m$, $w$), 
the comoving coordinates of all galaxies are computed, 
and the 2pCF is ready to be calculated.

\begin{figure*}
   \centering{
   \includegraphics[width=14cm]{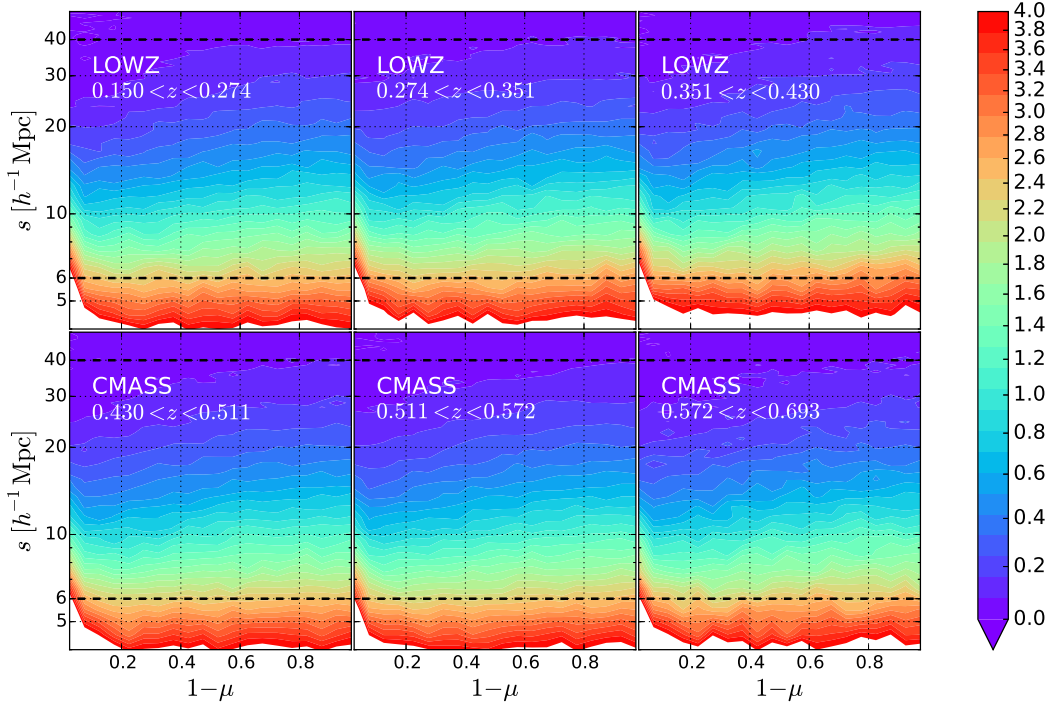}
   }
   \caption{
    \label{fig_2pcfcon} 
    2D contour map of measured $\xi$ as a function of $\mu$ and $s$, from the six redshift bins of LOWZ and CMASS samples 
      in the cosmology of $\Omega_m=0.31$ $\Lambda$CDM model.
    The black dashed lines mark the scales 6\ $h^{-1}$Mpc $\leq s\leq$ 40\ $h^{-1}$Mpc.
    The contour lines are not horizontal due to the effects of peculiar velocity.
    The FOG and Kaiser effects clearly manifest themselves through the tilting of contour 
     lines where $1-\mu \rightarrow 0$ and $1-\mu \gtrsim0.1$, respectively.
    The six contour maps have rather similar appearance, implying small redshift evolution of $\xi$.
   }
\end{figure*}

\subsection{Measuring the correlation function}

We adopt the Landy-Szalay estimator~\citep{1993ApJ...412...64L} to calculate the 2pCF,
\begin{equation}
\xi(s,\mu)=\frac{DD-2DR+RR}{RR}\ ,
\end{equation}
where $DD$ is the number of galaxy--galaxy pairs, 
$DR$ the number of galaxy-random pairs, 
and $RR$ is the number of random--random pairs, 
all separated by a distance defined by $s\pm\Delta s$ and $\mu\pm\Delta\mu$, 
where $s$ is the distance between the pair and $\mu=\cos(\theta)$, 
with $\theta$ being the angle between the line joining the pair of galaxies and the LOS direction to the target galaxy. 
This statistic captures the anisotropy of the clustering signal.

The random catalogue consists of unclustered points whose number density in redshift space mimics the radial selection function of the observational data. 
In an effort to reduce the statistical variance of the estimator we use 50 times as many random points as we have galaxies.
The galaxies and random points are weighted as described in Sec. \ref{sec:data}.

Figure \ref{fig_2pcfcon} shows the 2D contour of measured $\xi$ as a function of $\mu$ and $s$,
from the six redshift bins of LOWZ and CMASS samples 
in the cosmology of $\Omega_m=0.31$ $\Lambda$CDM model.
Due to the peculiar velocity effect, the contour lines are not horizontal.
The FOG \citep{FOG} and Kaiser \citep{Kaiser1987} effects 
clearly manifest themselves through the tilting of contour lines in regions of $\mu \rightarrow 1$ and $1-\mu \gtrsim0.1$, respectively.
A visual inspection of the contour maps from the six redshift bins 
reveals that they all have a similar appearance,
implying small redshift evolution of $\xi$.

\subsection{Probing the anisotropy through 2pCF}

The 2pCF is measured as a function of the separation $s$ and the angular direction $\mu$.
To probe the anisotropy we are more interested in the dependence of the 2pCF on $\mu$.
We follow the procedure of \cite{Li2015} and integrate the $\xi$ over the interval $s_{\rm max} \leq s \leq s_{\rm min}$.
We evaluate
\begin{equation}\label{eq:xideltas}
\xi_{\Delta s} (\mu) \equiv \int_{s_{\rm min}}^{s_{\rm max}} \xi (s,\mu)\ ds.
\end{equation}
The integration is limited at both small and large scales.
At small scales the value of $\xi$ is seriously affected by the FOG effect \citep{FOG}
which depends on the galaxies bias.
This may introduce a redshift evolution in $\xi_{\Delta s}(\mu)$ that is relatively difficult to model.
At large scales the measurement is dominated by noise due to poor statistics.
\cite{Li2015} found that $s_{\rm min}=6-10$ $h^{-1}$Mpc and $s_{\rm max}=40-70$ $h^{-1}$Mpc are reasonable choices 
which provide consistent, tight and unbiased constraints on cosmological parameters.
In this analysis we choose $s_{\rm min}=6$ $h^{-1}$Mpc and $s_{\rm max}=40$ $h^{-1}$Mpc.

The redshift evolution of the bias of observed galaxies leads to redshift evolution of the strength of clustering,
which is difficult to accurately model.
We mitigate this systematic uncertainty by relying on the shape of $\xi_{\Delta s}(\mu)$, rather than its amplitude,
\begin{equation}\label{eq:norm}
 \hat\xi_{\Delta s}(\mu) \equiv \frac{\xi_{\Delta s}(\mu)}{\int_{0}^{\mu_{\rm max}}\xi_{\Delta s}(\mu)\ d\mu}.
\end{equation}
We impose a cut $\mu<\mu_{\rm max}$ to reduce the fiber collision and FOG effects which are stronger toward the LOS ($\mu\rightarrow1$) direction.

\begin{figure*}
   \centering{
   \includegraphics[width=16cm]{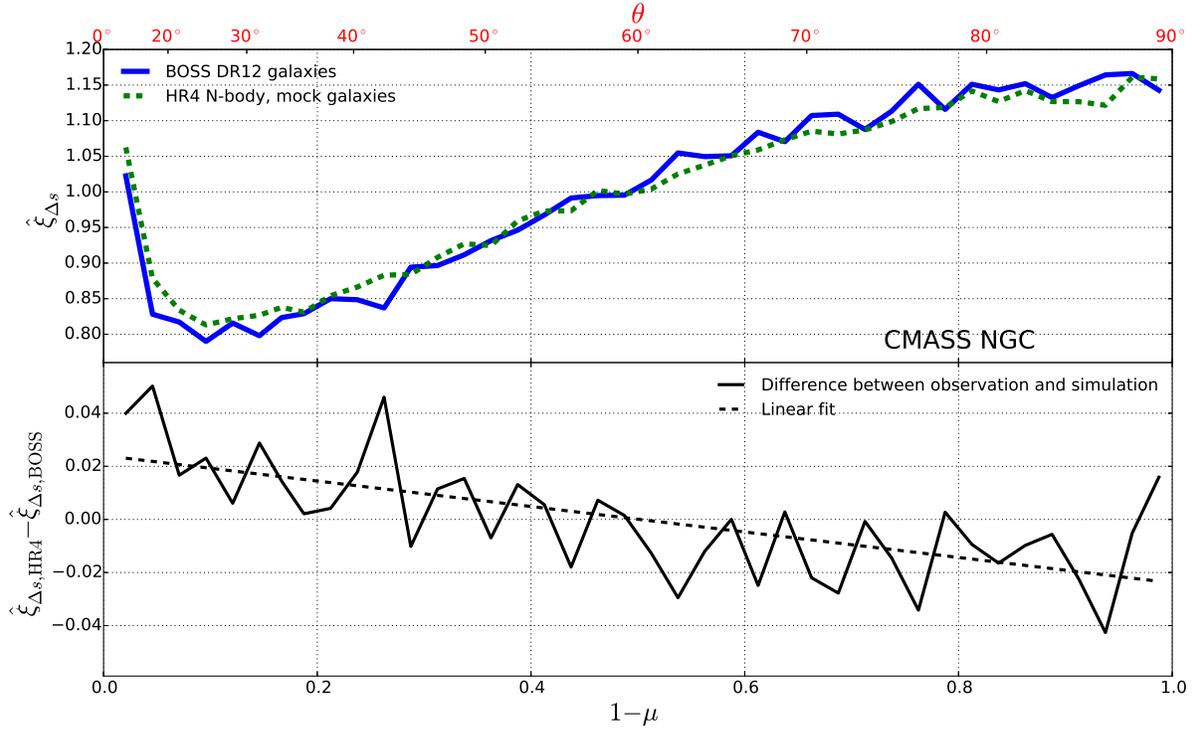}
   }
   \caption{\label{fig_datamock}
   $\hat\xi_{\Delta s}(\mu)$ measured from BOSS DR12 CMASS NGC sample 
    (consists of $\approx 565\,000$ galaxies at $0.15 < z < 0.693$)
    and one realization from the HR4 N-body simulation, in the WMAP5 cosmology.
   The lower panel shows the difference between the results of observational data and mock galaxies.
   $\hat\xi_{\Delta s}(\mu)$ are obtained by integrating $\xi (s,\mu)$ within the range {\rm 6\ $h^{-1}$Mpc}$\leq s\leq${\rm 40\ $h^{-1}$Mpc} 
      and normalizing the amplitude, 
    i.e., Eq. (\ref{eq:xideltas}) and Eq. (\ref{eq:norm}).
   To produce a clear view of the FOG effect, we split the angular range of $0.01\leq \mu \leq 1$ into as many as 40 bins.
   Using the HR4 mock galaxies, the enhancement near $\theta=0^\circ$ caused by the FOG effect
    and the tilt of the shape in $20^\circ\lesssim\theta\lesssim90^\circ$ 
    as a result of the large-scale flow,
    are all very well reproduced.
   This verifies the ability of our galaxies assignment method \citep{hong2016} 
    to reproduce the properties of galaxy distributions from large scale surveys.
   We use the HR4 mock galaxies to correct the systematics effects produced by the RSD.
   }
\end{figure*}

The clustering properties may be affected by various properties of the galaxy sample, 
such as, the mass, morphology, color, concentration. 
In our simulation, using the merger tree, 
we identify ``galaxies'', therefore we only use the galaxy mass building history to simulate BOSS galaxies. 
Therefore, it is necessary for us to test if our mock galaxies can accurately 
reproduce the $\hat\xi_{\Delta s}(\mu)$ of observed galaxies.


Figure \ref{fig_datamock}  compares the shape of $\hat\xi_{\Delta s}(\mu)$ measured from observational data and mock survey samples.
It is clear that $\hat \xi_{\Delta s}(\mu)$ from mock galaxies identified in the HR4 simulation (green dotted line) agrees well with the observation.
The enhancement near $\theta=0^\circ$ is caused by the FOG effect,
and the characteristic shape in $20^\circ\lesssim\theta\lesssim90^\circ$ 
produced from the large-scale flow
are all very well reproduced.
This result verifies the ability of our mock galaxies to reproduce the clustering properties of the observed galaxies.
The small overestimate (underestimate) of $\hat\xi_{\Delta s}$ at large (small) $\mu$ could be due to 
that our mocks galaxies are more massive than those in the observations.
We use the HR4 galaxy mocks to correct the systematics.

\begin{figure*}
   \centering{
   \includegraphics[width=16cm]{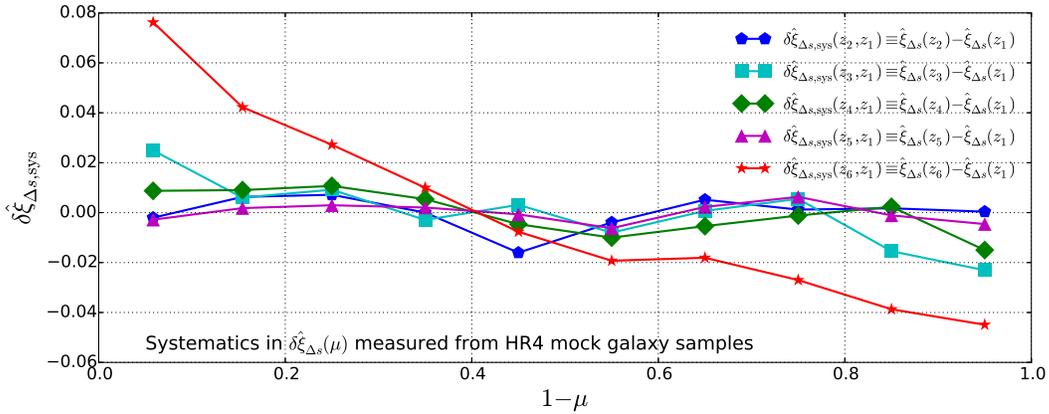}
   }
   \caption{\label{fig_sys}
   Systematics in $\delta \hat \xi_{\Delta s}$, measured from the HR4 mock galaxy samples.
   The redshift evolution of RSD effect and properties of samples can lead to non-zero values of $\delta \hat\xi_{\Delta s}(z_i,z_1)$.
   For the 1st to 5th redshift bins, $\delta\hat\xi_{\Delta s, \rm sys}(z_i,z_1)\lesssim0.02$,
    indicating a small redshift evolution of the RSD effect and properties of galaxies.
   For the 6th redshift bin, the values of $\delta\hat\xi_{\Delta s, \rm sys}(z_6,z_1)$ are relatively large,
   because galaxies in that highest redshift bin are significantly more massive than those at lower redshifts.
   See Sec. \ref{sec:syscor} for details.
   }
\end{figure*}

We divide the full angular range $0\leq\mu\leq\mu_{\rm max}$ into $n_{\mu}$ bins and measure its value in each bin.
Since we are free to choose $\mu_{\rm max}$ and $n_{\mu}$,
they are varied to optimize the S/N of our results.
This topic will be discussed in Sec. \ref{sec:binningscheme}.

\subsection{Characterizing the redshift evolution}

As shown in Figure \ref{fig_nbar}, we split the BOSS DR12 galaxies into six redshift bins, three in LOWZ and three in CMASS.
To study the redshift evolution of the clustering anisotropy 
we use the {\it first redshift bin} as the reference and compare the measurements in other bins with that in the first.
We define
\begin{equation} \label{eq:deltahatxi}
\delta \hat\xi_{\Delta s}(z_i,z_1,\mu_j)\ \equiv\ \hat\xi_{\Delta s}(z_i,\mu_j) - \hat\xi_{\Delta s}(z_1,\mu_j)
\end{equation}
where $\hat\xi_{\Delta s}(z_i,\mu_j)$ is $\hat\xi_{\Delta s}$ measured in the $i$th redshift bin and $j$th $\mu$ bin,
where $1\leq i \leq 6$ and $1\leq j \leq n_{\mu}$.
To characterize the shape of the curve well $n_\mu \gtrsim5$ is required.

\subsection{Correction for systematics}\label{sec:syscor}

Other than the AP effect, there are additional effects 
which may produce redshift-dependent anisotropy and affect the results. 

The observational artifacts, such as fiber collisions, redshift failures, 
and the non-cosmological density fluctuations induced by stellar density and seeing,
are accounted for in the galaxy weights \citep{Reidetal:2016}.
Fiber collisions and redshift failures may affect the value of $\hat\xi(\mu)$ in the region close to LOS;
we abandon the angular region of $1-\mu<0.01$, 
to avoid possible systematics (see Appendix \ref{sec:RBtest} for more discussion). 

The non-contiguous NGC and SGC are less well cross-calibrated with respect to each other 
than they are internally calibrated \citep{Schlafly2010,SF2011,Parejko2013}.
We construct the NGC and SGC mock surveys separately to avoid possible systematics.
The 2pCF analysis is also carried out for the NGC and SGC independently.
The result should be robust as long as each catalogue is well calibrated internally.



The apparent anisotropy introduced by RSD is, 
although greatly reduced by focusing on the redshift evolution, 
still the most significant systematic effect.

We estimate the value of $\delta \hat\xi_{\Delta s}$ from the systematic effects and subtract their contribution
(hereafter $\delta\hat\xi_{\Delta s, \rm sys}$) from the total variation.
The quantity $\delta\hat\xi_{\Delta s, \rm sys}$ is estimated from the HR4 mock galaxies.
The mock survey sample imitates the SDSS BOSS sample by mimicking
the survey as close as possible and includes past light cone effects.
The observational systematics such as the RSD, survey geometry, and shot noise
are included in the exactly same way as the observation.
The peculiar velocity perturbs the observed redshift through the relation
\begin{equation}\label{eq:zvpeu}
\Delta z = (1+z) \frac{v_{{\rm LOS}}}{c},
\end{equation}
where $v_{\rm LOS}$ is the LOS component of the peculiar velocity of galaxies.
The redshift evolution of galaxy peculiar velocities, 
resulting from growth of structure,
causes the anisotropy produced by RSD to have a small redshift evolution; 
this is the main source of systematic uncertainty in our results. 

We take the HR4 mock galaxy samples and compute $r(z)$ of galaxies in the cosmology under which the simulation is based.
In this case there is no AP effect. 
Thus, the measured $\delta \hat\xi_{\Delta s}$ are the redshift evolution purely created by systematics effects.
They are adopted as the estimation of $\delta\hat\xi_{\Delta s, \rm sys}$,
and the results are illustrated in Figure \ref{fig_sys}. 

For the 1st to 5th redshift bins, $\delta\hat\xi_{\Delta s, \rm sys}(z_i,z_1)\lesssim0.02$,
indicating a small redshift evolution of the RSD effect and properties of galaxies.
The only exception is the 6th redshift bin where the values of $\delta\hat\xi_{\Delta s, \rm sys}(z_6,z_1)$ are relatively large.
The reason for the large values is that the galaxies in that highest redshift bin are significantly more massive than those at lower redshifts,
so the measured high redshift $\hat\xi_{\Delta s}(\mu)$ has larger (smaller) values at $\mu\rightarrow1$ ($\mu\rightarrow0$) 
compared with the others
(an investigation of the dependence of $\hat\xi_{\Delta s}(\mu)$ on galaxy mass is provided in Sec. \ref{sec:caveats}).


\subsection{The caveats}\label{sec:caveats}

\cite{Li2014,Li2015} found the RSD effect exhibits a small redshift dependence of $\hat \xi_{\Delta s}(\mu)$, 
mainly due to the structure growth and the selection effect
(different galaxy bias at different redshifts).
In this analysis we use the mock galaxy sample from HR4 to correct this systematics.
The galaxy assignment scheme of \cite{hong2016} applied to HR4 is very successful in modeling both the 
large scale Kaiser effect and the small scale FOG effect in nonlinear regions.

There are two possible caveats in our procedure of the modeling of the RSD effect.

1) The RSD effect is estimated from mock survey samples created in a particular cosmology, 
i.e., the $\Omega_m=0.26$ $\Lambda$CDM model. 
If this adopted cosmology is different from the truth, then there could be a systematic bias in the estimation. 
We believe that this will not seriously affect our cosmological constraints.
\cite{Li2014} shows that the redshift dependence of RSD is not sensitive to cosmological parameters.
Also, the cosmologies adopted in simulations are consistent with our best-fit cosmological parameters within 1$\sigma$,
therefore our inferred cosmological constraints should be fairly accurate.
In a future analysis,
we will estimate the redshift evolution of the RSD effect from a set of cosmological simulations 
covering the relevant part of the parameter space.
This approach will remove the remaining uncertainty associated with the RSD effect, 
which is already a minor effect in our analysis.

\begin{figure*}
   \centering{
   \includegraphics[width=14cm]{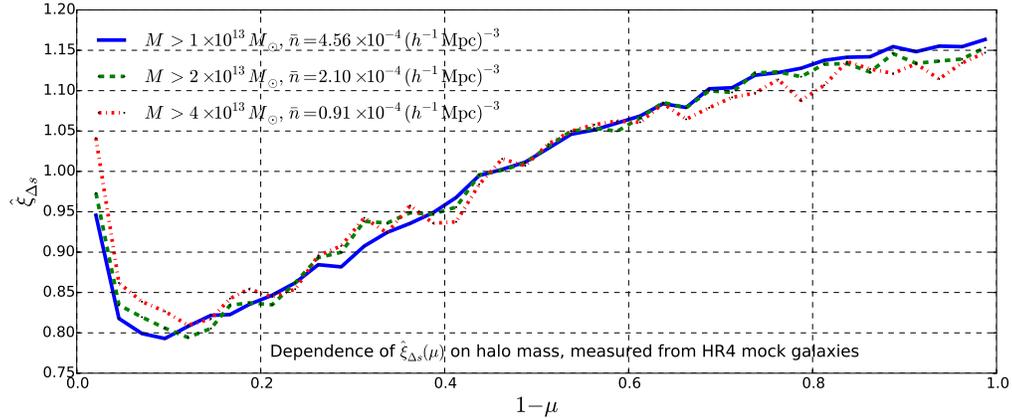}
   }
   \caption{
    \label{fig_2pcf_masscut} 
    $\hat \xi_{\Delta s}(\mu)$ measured from a small HR4 mock galaxy sample with different minimal mass cuts.
    The mock galaxies are taken from the $z=0$ snapshot data within the radius $r<600$ $h^{-1}$Mpc.
    Only small variations of $\hat\xi_{\Delta s}(\mu)$ are produced when changing the minimal mass cuts.
    Higher mass cuts result in larger (smaller) $\hat\xi_{\Delta s}(\mu)$ at $\mu\rightarrow1$ ($\mu\rightarrow0$).
   }
\end{figure*}

2) The selection effect, i.e., the evolution of galaxy bias with redshifts, 
can introduce redshift evolution in the clustering properties of the observed galaxies.
In our analysis the amplitude of the 2pCF is normalized and only its angular information, 
the function $\hat \xi_{\Delta s}(\mu)$, is used.
This function is rather insensitive to the galaxy bias,
which mainly affects the strength of clustering.

As a test, Figure \ref{fig_2pcf_masscut} shows $\hat \xi_{\Delta s}(\mu)$ measured from a small HR4 galaxy sample with different minimal mass cuts.
The mock galaxies are taken from the $z=0$ snapshot data within the radius $r<600$ $h^{-1}$Mpc.
Applying the minimal mass cuts of $1,2,4\times 10^{13} h^{-1} M_{\odot}$, 
we created three sets of subsamples with number density of 
$\bar n=4.56,\ 2.10,\ 0.91 \times 10^{-4} ( h^{-1} \rm Mpc)^{-3}$, 
which roughly covers the scatter of the number density of BOSS DR12 galaxies at $0.15<z<0.7$
\footnote{The BOSS LOWZ and CMASS galaxies reside in massive haloes 
with a mean halo mass of $5.2 \times 10^{13} h^{-1} M_{\odot}$ and $2.6 \times 10^{13} h^{-1} M_{\odot}$ \citep{Parejko2013,White2011,Reidetal:2016},
respectively.
For CMASS galaxies, when the redshift changes from $z=0.43$ to $0.7$,
the mean stellar mass varies from $10^{11.6} {M_{\odot}}$ to $10^{11.9} {M_{\odot}}$ \citep{CMASSLSS2014}.}.

For subsamples with higher mass cuts the $\hat\xi_{\Delta s}(\mu)$
has larger (smaller) values at $\mu\rightarrow1$ ($\mu\rightarrow0$).
More massive samples result in less tilted $\hat\xi_{\Delta s}(\mu)$ in the region of $1-\mu\gtrsim0.1$
\footnote{This phenomenon is understandable.
The tilt of $\hat\xi_{\Delta s}(\mu)$ is related to the RSD effect, 
and also the overall amplitude of the 2pCF (the denominator of Eq. (\ref{eq:norm})).
The slope should be roughly proportional to $(v/b_g)^2$, 
where the peculiar velocity term $v^2$ denotes the effect of RSD, 
and the galaxy bias term $b_g^2$ represents the amplitude of the 2pCF. 
For the more massive sample, $b_g$ is much larger while $v$ is still close to the peculiar velocity of dark matter field,
therefore the slope is smaller.
}.
This explains the relative large value of $\hat \xi_{\Delta s,\rm sys}(z_6, z_1)$.
In particular, comparing the subsamples with mass cuts $4\times 10^{13}h^{-1} M_{\odot}$ and $1\times 10^{13}h^{-1} M_{\odot}$, 
we find the red dotted curve is higher (lower) than the blue solid line at $\mu\rightarrow1$ ($\mu\rightarrow0$),
with a difference of $\approx$0.1 (0.05),
consistent with the value of $\hat \xi_{\Delta s,\rm sys}(z_6, z_1)$ shown in Figure \ref{fig_sys}.

In addition, this result also explains the small discrepancy between the $\hat\xi_{\Delta s}(\mu)$ measured from the 
observational data and the HR4 simulations (Figure \ref{fig_datamock}).
The mock galaxies could be systematically more massive than the observed ones.
These systematics could be most significant in the 6th redshift bin
where mock galaxies are most massive, 
leading to possible overestimation of $\delta\hat\xi_{\Delta s,{\rm sys}}$.
We discuss the impact of this effect in Sec. 6.1.

$\hat\xi$ is much less affected by the galaxy bias compared with the amplitude of $\xi$,
which is enhanced by 40\% and 100\% when increasing the mass cut from $1\times 10^{13}h^{-1} M_{\odot}$ to $2,4\times 10^{13}h^{-1} M_{\odot}$ .




\subsection{$\chi^2$ function}\label{sec:likelihood}

We define a $\chi^2$ function to quantify the redshift evolution of clustering anisotropy
\begin{equation}\label{eq:chisq1}
\chi^2\equiv \sum_{i=2}^{6} \sum_{j_1=1}^{n_{\mu}} \sum_{j_2=1}^{n_{\mu}} {\bf p}(z_i,\mu_{j_1}) ({\bf Cov}_{i}^{-1})_{j_1,j_2}  {\bf p}(z_i,\mu_{j_2}),
\end{equation}
where ${\bf p}(z_i,\mu_{j})$ is the redshift evolution of clustering, 
$\hat \xi_{\Delta s}$, with systematic effects subtracted
\begin{eqnarray}\label{eq:bfp}
 {\bf p}(z_i,\mu_{j}) \equiv&\ \delta \hat\xi_{\Delta s}(z_i,z_1,\mu_j) - \delta \hat\xi_{\Delta s, \rm sys}(z_i,z_1,\mu_j)
\end{eqnarray}
${\bf Cov}_i$ is the covariance matrix estimated from the mock surveys. 

The covariance matrix inferred from a finite number of 
Monte Carlo realizations 
is always a biased estimate of the true matrix \citep{Hartlap}.
This bias can be corrected by rescaling the inverse covariance matrix as 
\begin{equation}
 {\bf Cov}^{-1}_{ij,\rm Hartlap} = \frac{N_s - n_{\mu}-2}{N_s-1} {\bf Cov}^{-1}_{ij},
\end{equation}
where $N_s$ is the number of mocks used in covariance estimation.
In the case when using the 2\,000 MultiDark-Patchy mocks, 
the rescaling is only 1.008, 1.013, 1.018 if adopting $n_{\mu}=20,\ 30, \ 40$.

Also, the error of covariance matrix propagates to the 
error on the estimated parameters, 
leading to scattering in the inferred constraints \citep{Percival2014}.
Fortunately, this too can be easily corrected by multiplying the likelihood by a factor of
\begin{equation}
m_1 = \frac{1+B(n_\mu - n_{p})}{1+A+B(n_p +1)},
\end{equation}
where $n_p$ is the number of parameters , and $A,B$ take the forms of
\begin{eqnarray}
A = \frac{2}{(N_s-n_\mu-1)(N_s-n_\mu-4)},\\
B = \frac{N_s-n_\mu-2}{(N_s-n_\mu-1)(N_s-n_\mu-4)}.
\end{eqnarray}
We find $m_1=$1.011, 1.016, 1.021 for $n_{\mu}=$20, 30, 40, if using the 2\,000 MultiDark-Patchy mocks.

\subsection{Averaging a set of replicate measurements to increase the S/N}\label{sec:binningscheme}

The angular cut, $\mu_{\rm max}$, and number of bins, $n_\mu$, is chosen freely.
Different choices for these parameters yield slightly different results and different statistical uncertainties.
To suppress the statistical noise we adopt a large number of binning schemes and average their $\chi^2$s.

A value $\mu_{\rm max} = 0.99$ is sufficient to remove the fiber collision effect.
We have checked that our results are statistically robust against the choice of $\mu_{\rm max}$.

A larger $n_\mu$ tightens the constraint 
in the cost of more noise, 
while $n_\mu$ should be smaller than the number of mock samples used for covariance estimation.
We find $n_\mu \gtrsim 5$ gives relative tight constraints, 
and $n_\mu = 40$ is the limit we can reach given the size of the sample and the number of mock samples.

We run through the range $\mu_{\rm max} = 0.85, 0.86, ... 0.99$ at steps of 0.01 and $n_\mu =  6, ... 40$, in total 525 binning schemes.
We compute $\chi^2$ according to Equation (\ref{eq:chisq1}) for these choices and take an average of $\chi^2$s from a number of schemes.
This approach suppresses the statistical noise quite effectively. 



\begin{figure*}
   \centering{
   \includegraphics[width=18cm]{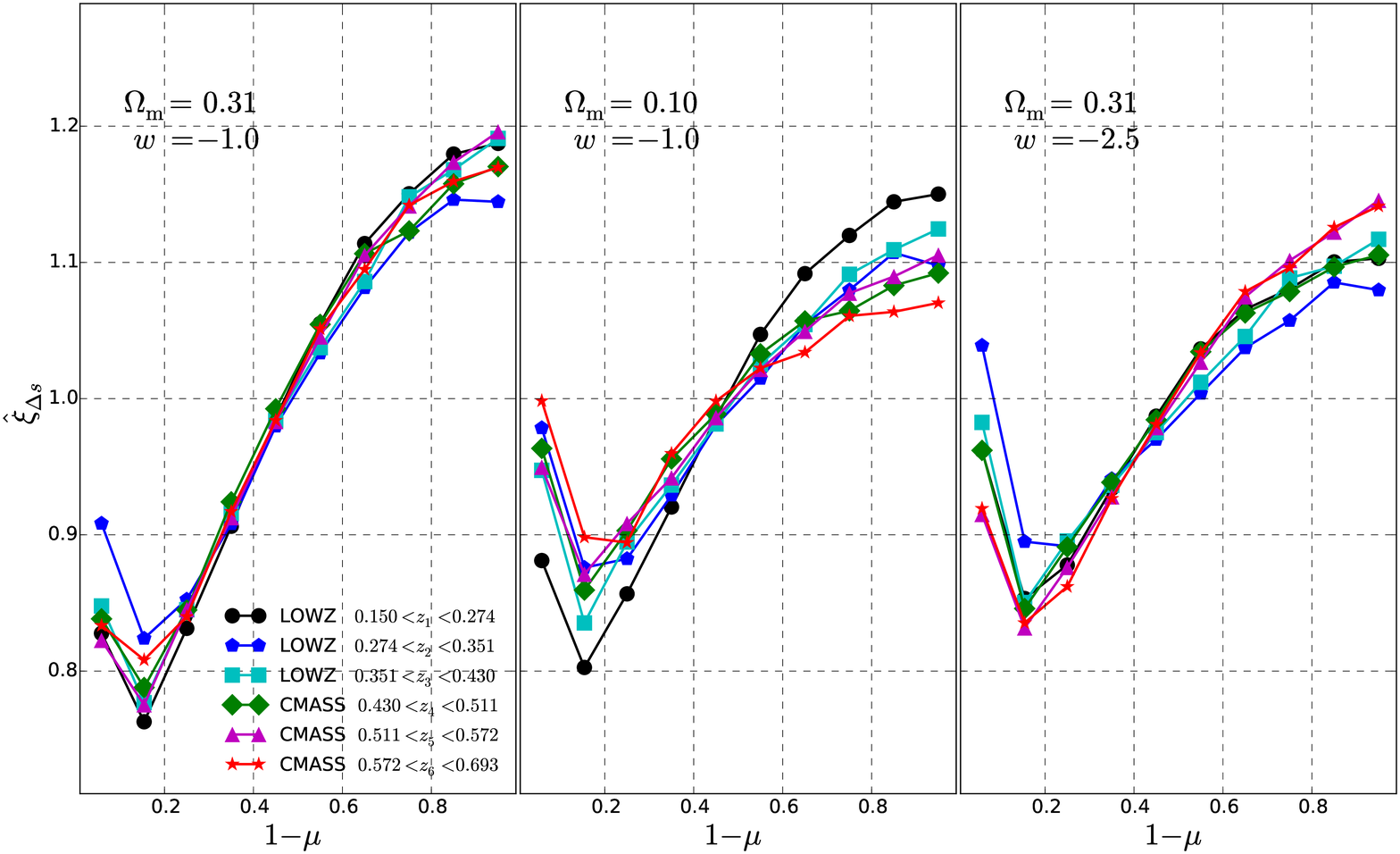}
   \includegraphics[width=18cm]{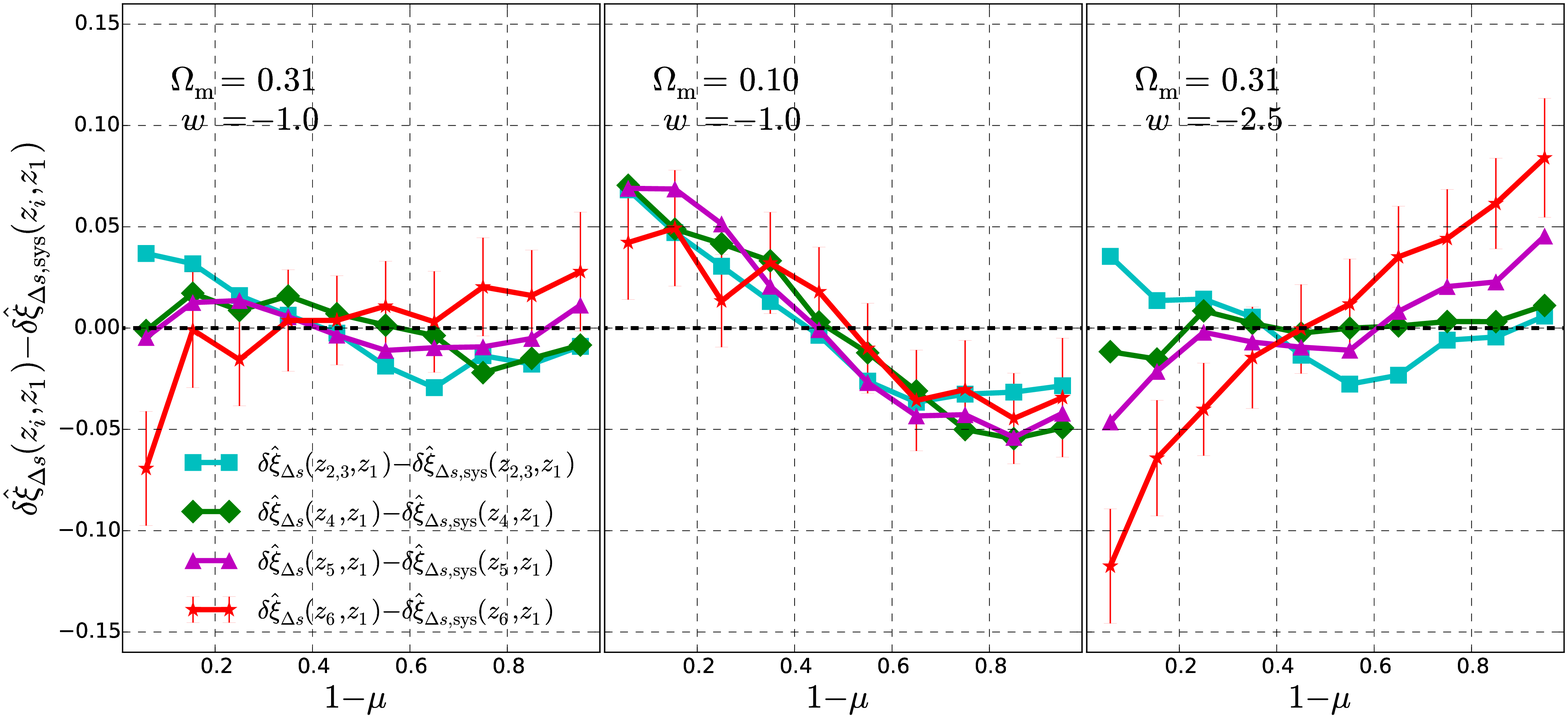}
   }
   \caption{\label{fig_TpCF}
   Upper panels: $\hat \xi_{\Delta s}(\mu)$ measured for the BOSS DR12 galaxies,
   which are divided into six redshift bins.
   Measurements from the three bins in LOWZ are indicated in solid lines, 
   while those from CMASS are the dashed lines.
   We present the measurements in three cosmologies,
   the $\Omega_m=0.31$ $\Lambda$CDM (left), the $\Omega_m=0.1$ $\Lambda$CDM (middle),
   and a cosmology with $\Omega_m=0.31$, $w=-2.5$ (right).   
   The angular dependence of 2pCF is measured in 10 bins within the range $0.01\leq1-\mu\leq1$.
   In the middle and right panels, the redshift dependence of AP distortions leads to clear redshift evolution of $\hat \xi_{\Delta s}(\mu)$.
   Lower panels: The redshift evolution of the $\hat\xi_{\Delta s}$ after systematics correction,  
   $\delta\hat \xi_{\Delta s}(z_i,z_1)-\delta\hat \xi_{\Delta s,\rm sys}(z_i,z_1)$
   (for details of their definitions, see Eq. (\ref{eq:deltahatxi}) and Sec. \ref{sec:syscor}).
   For clarity we combine the curves of $z_2$ and $z_3$, which fluctuate
   due to the relatively small sample sizes.
   The statistical significance is indicated by the 1$\sigma$ error bars of $\delta\hat \xi_{\Delta s}(z_6,z_1)-\delta\hat \xi_{\Delta s,\rm sys}(z_6,z_1)$.
   The curves are statistically consistent with 0 in the $\Omega_m=0.31$ $\Lambda$CDM cosmology,
    while they deviate from 0 at significant confidence levels in the $\Omega_m=0.1$ cosmology and the $w=-2.5$ cosmology.
   Applying the likelihood analysis described in Sec. \ref{sec:likelihood},
    the two latter models disfavored at 5.3$\sigma$ and 7.5$\sigma$ CL compared to the $\Omega_m=0.31$ $\Lambda$CDM cosmology.
   }
\end{figure*}

\section{Result}

We apply our method to the BOSS DR12 CMASS and LOWZ galaxies.
We present the results in this section.

\subsection{Redshift evolution in wrong cosmologies}\label{sec:redevolvxi}


The upper panels of Figure \ref{fig_TpCF} present the $\hat \xi_{\Delta s}$ measured for BOSS galaxies in three different cosmologies.
In the left panel we adopt the Planck cosmology, i.e., $\Lambda$CDM with $\Omega_m=0.31$;
In the other two panels, we choose two sets of parameters ---
the $\Omega_m=0.1$ $\Lambda$CDM cosmology (middle) and the cosmology with $\Omega_m=0.31$ and $w=-2.5$ (right).
We split the BOSS DR12 galaxies into six redshift bins,
construct their 3D distribution in these three cosmologies
and then measure $\hat \xi_{\Delta s}(\mu)$ according to the procedure of the last section.

For a sample of galaxies with homogeneous, isotropic spatial distribution, 
the measured $\hat \xi_{\Delta s}(\mu)$ is statistically uniform.
The distortion in $\hat \xi_{\Delta s}(\mu)$ introduced by RSD is large.
There are two distinct features in this distortion.
At $0\lesssim 1-\mu \lesssim 0.1$, $\hat \xi_{\Delta s}(\mu)$ turns up as $\mu\rightarrow1$
due to the large FOG effect.
In other directions, the anisotropic clustering produced by the large scale flow creates a monotonic angle dependence of $\hat\xi_{\Delta s}(\mu)$. 
These patterns are evident in all cosmologies.


The additional angular dependence of $\hat \xi_{\Delta s}(\mu)$ introduced by AP is not as strong as RSD;
However, it is still visible. 
In the middle and right panels, 
the choice of two incorrect cosmologies results in an enhancement of  $\hat \xi_{\Delta s}(\mu)$ due to the 
apparent stretch of non-linear structures in the LOS direction
\footnote{{\it Stretch of structure enhances the value of $\xi_{\Delta s}$}.
On relatively small scales, the strong clustering produces large values of $\xi$.
The apparent stretch of structure means the strong clustering on small scales are, apparently, shifted to larger scales:
at some fixed scale, we measured larger $\xi$ if there is apparent stretch.
The value of $\xi_{\Delta s}$, which is the integral of $\xi$ within fixed range of $s$, is also enhanced.}.
This effect is evident when we compare  $\hat \xi_{\Delta s}(\mu)$ in these two panels to that in the left panel.

More importantly, in incorrect cosmologies the redshift dependence of the AP effect results in a redshift evolution of $\hat \xi_{\Delta s}$.
This unique feature makes the AP effect detectable even with the existence of the large RSD effect. 
When $\Omega_m=0.1$ and $w=-1$ are adopted, 
the LOS stretch of structure becomes stronger at higher redshift (see Figure \ref{fig_xy});
as a result, the enhancement of $\hat \xi_{\Delta s}$ along the LOS 
becomes more significant at higher redshifts.
The opposite behavior is seen for the choice of $\Omega_m=0.31$ and $w=-2.5$.
However, we do not see such obvious redshift evolution in the Planck cosmology, 
 indicating that the cosmological parameters of the Planck cosmology are close to the correct ones.

The lower panels of Figure \ref{fig_TpCF} show the redshift evolution of $\hat\xi_{\Delta s}$ after systematics correction, 
i.e., the quantity $\delta\hat \xi_{\Delta s}-\delta\hat \xi_{\Delta s,\rm sys}$.
The result is statistically consistent to 0 in Planck cosmology -- 
In the four curves, almost all points are consistent with 0 at $\lesssim1\sigma$ confidence level (CL).
The only exception is the leftmost point of $\delta\hat \xi_{\Delta s}(z_6,z_1)-\delta\hat \xi_{\Delta s,\rm sys}(z_6,z_1)$,
 which is negative at $\sim$2 $\sigma$ CL.
This result could be due to the overestimation of $\delta\hat\xi_{\Delta s, \rm sys}(z_6,z_1)$ as discussed in Sec. \ref{sec:syscor}.
This behavior will not affect our derived cosmological constraints:
Dropping this measurement by imposing a cut $\mu<0.9$
  does not shift the constrained on $\Omega_m$-$w$ (the pink contour of Figure \ref{fig_contours}).

In the $\Omega_m=0.1$ cosmology and the $w=-2.5$ cosmologies, however,
the measured $\delta\hat \xi_{\Delta s}-\delta\hat \xi_{\Delta s,\rm sys}$ deviates 
from zero at significant confidence levels.
Applying the likelihood analysis described in Sec. \ref{sec:likelihood},
 they are disfavored at 5.3$\sigma$ and 7.5$\sigma$ CL compared to the Planck cosmology.
Since these two cosmologies deviates from the $\Omega_m=0.31$ $\Lambda$CDM cosmology with $\Delta \Omega_m = 0.21$ and $\Delta w = 1.5$,
roughly speaking, we expect our method able to constrain $\Omega_m$ and $w$ with 1$\sigma$ uncertainties of 0.04 and 0.2, respectively.


\subsection{Cosmological constraint}\label{sec:constraint}

We constrain $\Omega_m$ and $w$ through Bayesian analysis \citep{Bayesian},
which derives the probability distribution function (PDF) of some parameters $\bf \theta$ (=($\Omega_m,w$) in this paper)
given observational data $\bf D$, 
according to Bayes' theorem:
\begin{equation}
 P({\bf \theta}|{\bf D}) = \frac{P({\bf \theta})P({\bf D}|{\bf \theta})}{m({\bf D})}.
\end{equation}
Here $P(\bf \theta)$, the prior distribution of the parameters,
contains all the information about the parameters known from substantive knowledge 
and expert opinion {\it before} observing the data.
The marginal PDF of $\bf D$, 
$m({\bf D})=\int P({\bf D|\theta})P({\bf \theta})d{\bf \theta}$, 
is a normalization constant independent of $\theta$.
All the information about the parameter $\bf \theta$ that stems from the experiment
is contained in the function $P({\bf D}|{\bf \theta})$, 
the conditional PDF of the observation $\bf D$ given the value of parameter.

\begin{figure*}
   \centering{
   \includegraphics[width=16cm]{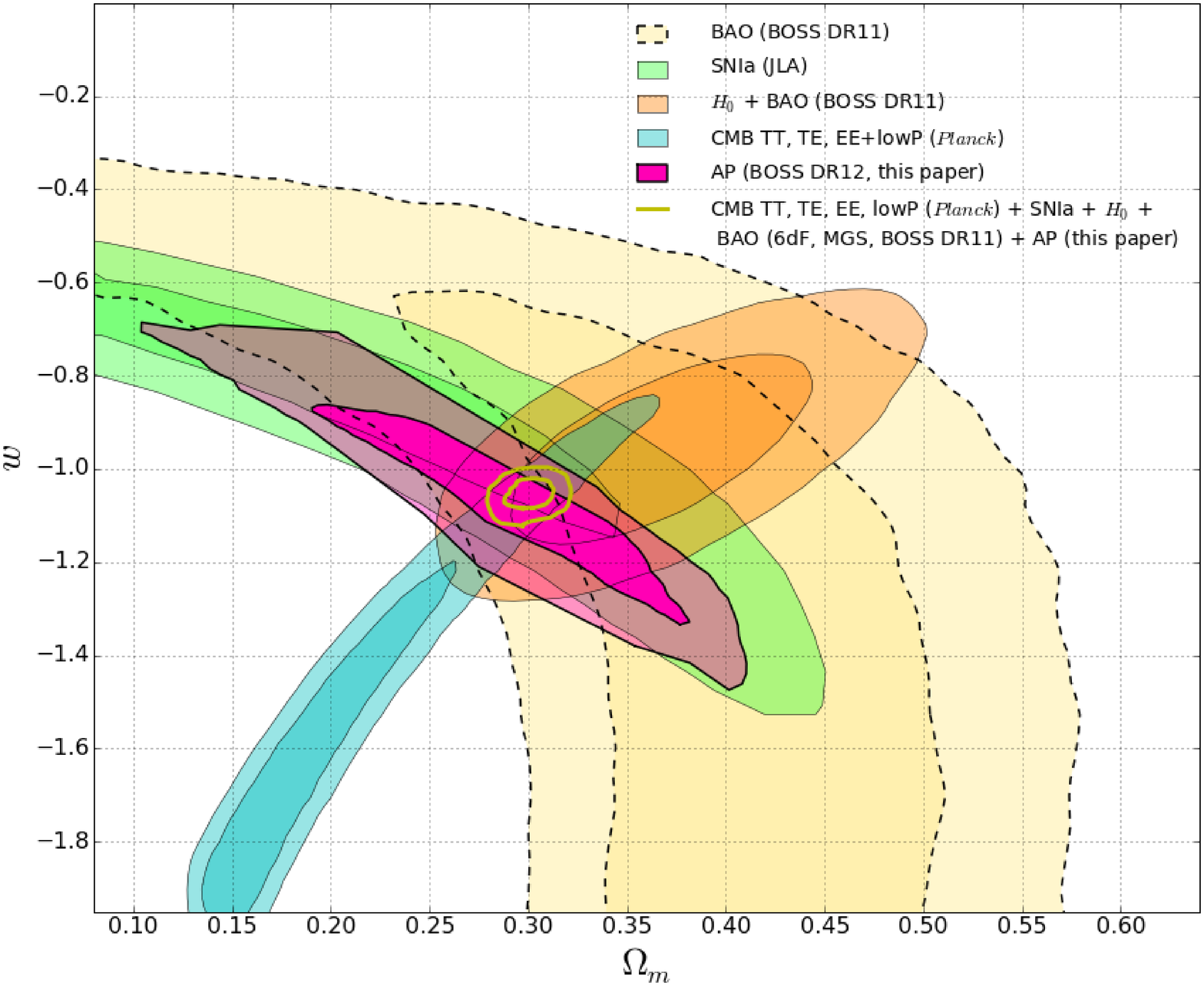}
   }
   \caption{\label{fig_contours}
   Likelihood contours (68.3\%, 95.4\%) in the $\Omega_m-w$ plane from our method and other cosmological probes.
   Using the BOSS DR12 galaxies within the redshift range $0.15< z< 0.693$, our method achieves tight cosmological constraints on $\Omega_m$ and $w$.
   The 2,000 MultiDark-Patchy mocks are adopted to estimate the covariance matrix in our method.
   Constraints from various probes are consistent with each other.
   See Sec. \ref{sec:constraint} for details.
   }
\end{figure*}

In this analysis we simply assume flat priors for $\Omega_m$ and $w$,
and approximate $P({\bf D}|{\bf \theta})$ by a likelihood function $\mathcal{L}$
satisfying  $-2 \ln \mathcal{L}=\chi^2$;
the PDF of $\theta$ derived from our AP method takes the form
\begin{equation}
 P({\bf \theta}|{\bf D}) \propto \mathcal{L} \propto \exp\left[-\frac{\chi^2}{2}\right].
\end{equation}
We use the {\texttt {COSMOMC}} software \citep{LB2002}
to obtain the Markov Chain Monte Carlo (MCMC) samples of $\theta$ following the PDF of $P({\bf \theta}|{\bf D})$.
Constraints on $\Omega_m$ and $w$ are derived from these samples.

We utilize the 2\,000 MultiDark-Patchy mocks to compute the covariance matrix, 
and take the averaged value of the $\chi^2$ map from
the 80 binning schemes with $20\leq n_\mu \leq 35$ 
and $\mu_{\rm max}\leq 0.95, ..., 0.99$
\footnote{The result is not sensitive to the choice of the binning scheme.}. 
The 68\% and 95\% likelihood contours of $\Omega_m$ and $w$ 
obtained from this analysis are shown in Figure \ref{fig_contours} (pink areas).
Our AP method yields tight constraints on $\Omega_m$ and $w$ .
The mean values and standard deviations are
\begin{equation}
 \Omega_m=0.290 \pm 0.053,\          \ w = -1.07 \pm 0.15. 
\end{equation}
This result is consistent with the Planck $\Lambda$CDM cosmology within 1$\sigma$ \citep{Planck2015}.

We find the expected negative degeneracy between $\Omega_m$ and $w$.
At low redshift, increasing $\Omega_m$ and increasing $w$ have a similar effect on the cosmic expansion.
When one parameter is increased the other parameter can be decreased to counteract the AP effect
\footnote{At high redshift the direction of degeneracy completely changes. 
There the effect of $w$ is less important and the cosmic expansion is mainly governed by $\Omega_m$.
This behavior is why \cite{Li2015} obtained a positive degeneracy between $\Omega_m$ and $w$ from mock surveys 
spanning a wide range of $0< z < 1.5$.
The degeneracy direction between $\Omega_m$ and $w$ depends on the redshift range of the sample used in the analysis.}.

For comparison, additional likelihood contours from other cosmological probes are displayed in Figure \ref{fig_contours}:
the JLA SNIa sample (green area, \cite{JLA}),
the BAO measurement from BOSS DR11 galaxies (bright yellow, \cite{Anderson2013}),
a combination of the BOSS DR11 BAO and the Hubble Space Telescope measurement of $H_0=70.6\pm3.3$ (dark yellow, \cite{Riess2011,E14H0}),
and the full-mission Planck observations of temperature and polarization anisotropies, released in 2015 (blue, \cite{Planck2015}).

The SNIa contours occupy a region similar to ours, but provide much weaker constraints. 
The direction of degeneracy is also similar.
By measuring the apparent magnitudes of type Ia supernovae distributed at different redshifts, 
 cosmologists can infer the luminosity distance $d_L$ as a function of redshift.
The JLA sample has similar redshift coverage to the BOSS DR12 galaxies,
which could be the reason for the similarity between the contours of the SNIa and our AP method.


BAO itself can not effectively constrain $\Omega_m$ and $w$, 
since these parameters are highly degenerate with $H_0$ in determining the length scale.
Combining BAO with the $H_0$ measurements to break the degeneracy yields better constraint on the parameter space.
The direction of degeneracy of the combined constraint is roughly orthogonal to ours.

CMB measurements provide a powerful probe of the geometry of the Universe,
but can not effectively constrain $\Omega_m$ and $w$ due to the strong degeneracy of the parameters.
The CMB contour is almost orthogonal to the AP contour. 

Among all the individual cosmological probes, our method yields the most stringent 
constraint on $\Omega_m$ and $w$.
Our result is consistent with those of all other probes.

Assuming that the five different cosmological probes of CMB, BAO, SNIa, $H_0$ and our method are 
statistically independent,
we combine the results by simply multiplying their likelihoods,
resulting in the total likelihood function
\begin{equation}\label{eq:likelihood}
\mathcal{L}_{\rm total} = \mathcal{L}_{\rm CMB} \times \mathcal{L}_{\rm BAO} \times \mathcal{L}_{\rm SNIa}\times \mathcal{L}_{\rm H_0}
 \times \mathcal{L}_{\rm Our\ AP}.
\end{equation}
The Planck team has released the {\texttt {COSMOMC}} outputs of MCMC samples
using CMB+BAO+JLA+$H_0$
\footnote{http://pla.esac.esa.int/pla/\#cosmology; 
Their BAO datasets include three sets of measurements from SDSS DR11 \citep{Anderson2013}, 6dFGS \citep{6dFGS} and SDSS MGS \citep{MGS}}.
We modify their MCMC samples (by changing the values of sample weight and likelihood)
 to include the likelihood of our method,
creating samples following the PDF of Eq. (\ref{eq:likelihood}).
The joint constraint from all five cosmological probes is
\begin{equation}
 \Omega_m = 0.301 \pm 0.006,\ w=-1.054 \pm 0.025. 
\end{equation}
Interestingly, we find that the constraint is in tension with $w=-1$ at 2.1$\sigma$ CL,
However, the statistical evidence is not strong enough to claim deviation from the cosmological constant.

Our AP method plays an important role in strengthening the constraints on $\Omega_m$ and $w$.
As a comparison, 
a combination of CMB and BAO yields 
$\Omega_m = 0.306 \pm 0.013,\ w=-1.03 \pm 0.06$;
in the absence of our method, a combination of SNIa, CMB, BAO and $H_0$
yields $\Omega_m = 0.306 \pm 0.009,\ w=-1.03 \pm 0.04$ \citep{Planck2015}.


\section{Concluding Remarks}

We apply the methodology developed in \cite{Li2014,Li2015} to BOSS DR12 galaxies.
In LSS surveys, the observed galaxy distribution appears anisotropic due to two reasons:
the contamination of galaxy redshifts due to the galaxy peculiar velocities, known as the RSD effect, 
and the error in the distance due to inaccurate cosmological parameters. 
\cite{Li2014} reported that anisotropies produced by RSD effect are, although large,
close to uniform in magnitude over a large range of redshift,
while the degree of anisotropies introduced by AP varies with redshift. 
Thus we can use the redshift dependence of the anisotropic clustering of galaxies to constrain cosmological parameters without 
being much affected by RSD.

As in \cite{Li2015}, we investigate the redshift-dependence of clustering anisotropy by examining the 2pCF.
The 2pCF measured along different angular directions was characterized by the function $\hat \xi_{\Delta s}(\mu)$.
When the cosmological parameters governing the expansion history of the universe are incorrectly chosen,
the shape of this function evolves with redshift.
We split the DR12 galaxies into six redshift bins, measure the 2pCF in each redshift bin,
and search for the underlying true values of $\Omega_m$ and $w$ of our Universe 
by requiring minimal redshift evolution of $\hat \xi_{\Delta s}(\mu)$ across the six redshift bins.
We obtain tight constraints of 
 $\Omega_m=0.290 \pm 0.053,\          \ w = -1.07 \pm 0.15$ 
from our method alone.

The constraints on $\Omega_m$ and $w$ from our AP method are comparable with or tighter 
than the other cosmological probes of SNIa, CMB, BAO, and $H_0$.
For the direction of degeneracy, 
our method is similar to SNIa and orthogonal to CMB and BAO+$H_0$.
Combining the results of our method with those of other cosmological probes, 
we obtain tight constraints 
$\Omega_m = 0.301 \pm 0.006,\ w=-1.054 \pm 0.025$.

\subsection{Comparison with other LSS probes}

Our method uses the anisotropic galaxy on scales from 6 to 40 $h^{-1}$Mpc.
Constrains on cosmological parameters are obtained from the redshift dependence of $D_A(z) H(z)$.

As a comparison, the BAO method uses the BAO feature in the 
clustering of galaxies on scales of 100-150 $h^{-1}$Mpc 
created by the oscillation of the baryon-photon plasma in the early Universe.
Measuring the BAO feature in 1D or 2D yields measurements of $D_V$ or $D_A$ and $H$ at some representative redshifts.

The AP methods proposed to date, such as those using 
galaxy pairs and voids, measure the rate of geometric distortion and are sensitive to $D_A(z) H(z)$,
while our method uses its the redshift dependence.
These methods can be combined together to fully utilize the physics of AP test.
Also, reducing the RSD effect 
through the redshift dependence could be 
applicable to these methods.

The topology method proposed by \cite{topology} 
uses the redshift evolution of the volume effect and is sensitive to the quantity $D_A(z)^2 / H(z)$.
Combing this method with ours can yield separate constraints on $D_A$ and $H$ 
for the same observational sample.
Constraints from the other statistical measures,
such as the distribution function of size or richness of LSS,
can be also combined \citep{Park2012,Park2015}.

Recently, \cite{MS2016} developed a novel method 
constraining cosmological parameters based on the high-level similarity of the emission measure in the cluster outskirts.
In incorrect cosmologies, the emission measure from clusters exhibits redshift dependence.
Utilizing a sample of 320 galaxy clusters ($0.056<z<1.24$) observed with Chandra,
they achieve tight cosmological constraints comparable to ours.
The idea of this novel technique is to some extent similar to our method 
(seeking for the conservation of geometric quantity with redshift),
and could have promising future.

The above geometric methods can be combined with probes of RSD 
\citep{Guzzo2008,Blake2011a,Beutler2012,Reid2012b,Samushia2012,GM2016b,Li2016} 
to have a more complete study of LSS galaxy clustering.
See \cite{DHW2013} for a review of more LSS probes of dark energy.


\subsection{Room for improvements}

This paper is the first application of this redshift dependent AP test to observed LSS data.
There remains considerable opportunity for improving the analysis methodology,
e.g., the optimized schemes of redshift binning and 
optimized choices of the scales of clustering (the values $s_{\rm min}$ and $s_{\rm max}$).
In this paper, the 2pCF is characterized by $\hat \xi_{\Delta s}(\mu)$.
It should be more advantageous to use 
the redshift evolution of 2pCF in two dimensions, i.e., $\hat \xi(s,\mu)$, 
to capture the full information.
It is also necessary to combine the 
information in the higher-order
statistic beyond the two point functions.
As pointed out in Sec. \ref{sec:caveats},
more theoretical and numerical studies on the 
redshift evolution of the RSD effect 
will remove the remaining small 
uncertainties in our results.

To avoid the difficulty of modeling galaxy bias we dropped the information of strength of clustering
through normalizing the amplitude of $\xi$.
In the case that we have good knowledge of galaxy bias, one can utilize the amplitude of $\xi$ to probe the redshift dependence of volume effect 
and obtain much tighter constraints.
\cite{Li2015} showed that
the area of constraining regions in $\Omega_m$ and $w$ space from the redshift dependence of the volume effect is 3.5 times smaller 
than that from the redshift dependence of the AP effect alone.

In this analysis, by reducing the RSD effect, 
we are able to use the galaxy clustering down to 6 $h^{-1}$Mpc.
This is a major advance in extracting the cosmological information 
on small scales where galaxy clustering is strong
and there are a lot of independent structures.

To make sure that the derived cosmological constraint is robust, 
in the procedure of systematics correction, 
one should construct many mock surveys in which the RSD and other systematic effects are reliably modeled.

It would also be helpful to investigate the effect of galaxy properties.
For example, in case of a dense survey containing galaxies with various properties, 
one can split the galaxy sample into subsamples with different galaxy properties, 
derive cosmological constraints from these subsamples, 
and check the consistency of the results. 



\subsection{Synergies with future observations}

The constraining power of our method is proportional not only to the size, 
but also to the redshift coverage of the galaxy sample.
In this analysis we utilize 1\,133\,326 BOSS DR12 galaxies covering $0.15< z < 0.693$.
Future redshift surveys such as eBOSS \citep{eBOSS}, EUCLID \citep{EUCLID}, and DESI \citep{DESI}
will have larger sample sizes and wider redshift coverage.
Our method is expected to yield much tighter cosmological constraints
and can be applied to constrain wider classes of 
cosmological models when these data are available.

In this analysis we assume a flat Universe to constrain $\Omega_m$ and $w$.
Our method should be sensitive to all other cosmological parameters governing the cosmic expansion history,
e.g., the curvature of the Universe, 
and the other dark energy parameters.

Overall, the future of our method is extremely promising. 
With the progressive development of LSS experiments, 
we expect it to play an important role in deriving cosmological constraints from LSS surveys.

\section*{Acknowledgments}

We thank the Korea Institute for Advanced Study for providing computing resources (KIAS Center for Advanced Computation Linux Cluster System).
We would like to thank Ho Seong Hwang, Francisco-Shu Kitaura, Benjamin L'Huillier, Donghui Jeong, Teppei Okumura, Will Percival, Hyunmi Song 
and Yi Zheng for kind helps and helpful discussions.
This work was partially supported by the
Supercomputing Center/Korea Institute of Science and
Technology Information with supercomputing resources
including technical support (KSC-2013-G2-003).

Based on observations obtained with Planck (http://www.esa.int/Planck), 
an ESA science mission with instruments and contributions directly funded by 
ESA Member States, NASA, and Canada.

Funding for SDSS-III has been provided by the Alfred P. Sloan Foundation, the Participating Institutions, the
National Science Foundation, and the U.S. Department of Energy Office of Science. 
The SDSS-III web site is http://www.sdss3.org/. 
SDSS-III is managed by the Astrophysical Research Consortium for the Participating Institutions
of the SDSS-III Collaboration including the University of Arizona, the Brazilian Participation Group, Brookhaven
National Laboratory, Carnegie Mellon University, University of Florida, the French Participation Group, 
the German Participation Group, Harvard University, the Instituto de Astrofisica de Canarias, the Michigan State/Notre
Dame/JINA Participation Group, Johns Hopkins University, Lawrence Berkeley National Laboratory, Max Planck
Institute for Astrophysics, Max Planck Institute for Extraterrestrial Physics, New Mexico State University, New
York University, Ohio State University, Pennsylvania State
University, University of Portsmouth, Princeton University,
the Spanish Participation Group, University of Tokyo, University of Utah, Vanderbilt University, University of Virginia, 
University of Washington, and Yale University.

\appendix

\section{A. Robustness Tests}\label{sec:RBtest}

We conduct a series of tests to check the robustness of our results.

The cosmological constraints presented in Sec. \ref{sec:constraint} (hereafter the ``main results'') are derived using 
the following options (hereafter the ``default options''):
\begin{itemize}
 \item We adopt $s_{\rm min}=6 h^{-1} {\rm Mpc},\ s_{\rm max}=40 h^{-1} {\rm Mpc}$ in the integrated 2pCF
 ($\xi_{\Delta s} (\mu) \equiv \int_{s_{\rm min}}^{s_{\rm max}} \xi (s,\mu)\ ds$).
 \item The effects of systematics are estimated from the HR4 mock galaxies. 
 When constructing the mocks, to determine when a satellite galaxy is disrupted,
 we adopt the J08 model to calculate the satellite galaxy merger timescale.
 \item In total 2\,000 sets of MultiDark-Patchy mock surveys are used to construct the covariance matrix.
 \item We drop the angular region $1 - \mu < 0.01$, to mitigate the effects of fiber collisions and redshift failures.
\end{itemize}

\begin{figure*}
   \centering{
   \includegraphics[width=16cm]{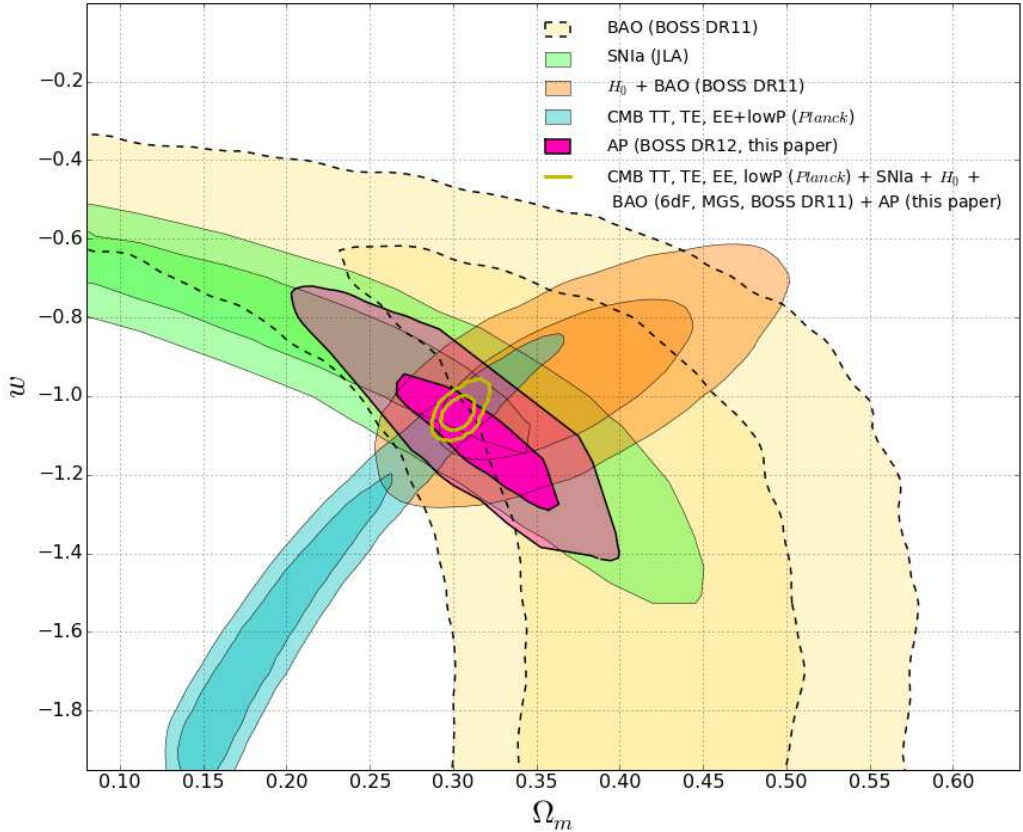}
   }
   \caption{\label{fig_contours_HR3}
   Likelihood contours (68.3\%, 95.4\%) in the $\Omega_m-w$ plane from our method and other cosmological probes.
   The 72 HR3 mock surveys are utilized to estimate the covariance matrix;
    the corrections of \cite{Hartlap} and \cite{Percival2014} are not applied.
   }
\end{figure*}

\begin{figure*}
   \centering{
   \includegraphics[width=14cm]{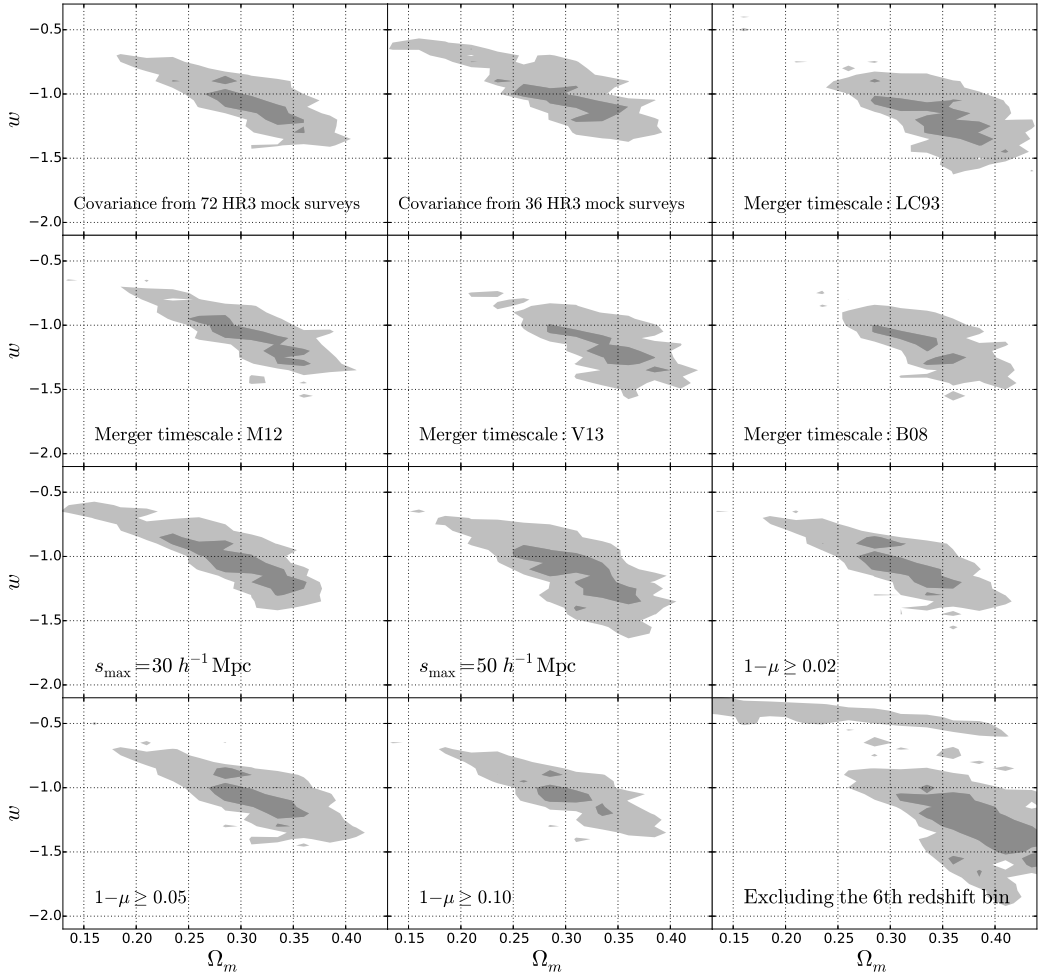}}
   \caption{ \label{fig_contour_RB}  
   Likelihood contours (68\%, 95.4\%) in the $\Omega_m$-$w$ plane, derived by applying our AP method to the BOSS DR12 galaxies.
    Different options are adopted to test the robustness (see Appendix \ref{sec:RBtest} for details). 
   }
\end{figure*}

We check whether the results are sensitive to these options.
We modify the above options one by one, 
derive the corresponding cosmological constraints, 
and check how much they deviate from the main results.
We present this information in Figure \ref{fig_contour_RB}. 

We use the 72 sets of HR3 PSB mock surveys to construct the covariance matrix,
and adopt a coarser grid with $\delta \Omega_m=0.025,\ \delta w=0.05$ when covering the parameter space.
This resolution is high enough for the purpose of testing our methodology.
We adopt the averaged $\chi^2$ from the 525 binning schemes
with $6\leq n_\mu \leq 40$ and $\mu_{\rm max} \geq 0.85, ..., 0.99$.

When using the HR3 mock surveys, we do not apply the correction factors of \cite{Hartlap} and \cite{Percival2014}
as they are not accurate when $n_\mu$ is not much smaller than $N_s$.
The inferred constraints could suffer from statistical scatter caused by the error and bias in the estimated covariance matrix.
The results presented in this section are therefore only for illustrative purposes.

\subsection{$Covariance\ Estimation$}

In this analysis we use a set of 2\,000 MultiDark-Patchy mock surveys to compute the covariance matrix.
We check whether the result could change if adopting the HR3 N-body mock surveys.

The cosmological constraints in case of adopting the covariance matrix estimated from HR3 mock surveys
and the {\it high resolution grid} is displayed in Figure \ref{fig_contours_HR3}.
The results are consistent with the main results derived using the 2\,000 MultiDark-Patchy mocks.
The constrained area shrunk, particularly towards small $\Omega_m$ and large $w$ values;
that does not much affect the combined cosmological constraints.

In case of using HR3 mocks, the constraint from our method alone is
\begin{equation}
 \Omega_m=0.314 \pm 0.038,\ \ w = -1.09 \pm 0.14,
\end{equation}
and the joint constraint from all five cosmological probes is
\begin{equation}
 \Omega_m = 0.304 \pm 0.007,\ w=-1.04 \pm 0.03.
\end{equation}
The difference between the above constraints and the results using the 2\,000 MultiDark-Patchy mocks is $\lesssim0.5\sigma$.

The first and second panels of the first row of of Figure \ref{fig_contour_RB} show  
the cosmological constraints from the {\it coarse grid}, 
using the covariance matrix estimated from 72 and 36 HR3 mock surveys, respectively.
They are consistent with the results displayed in Figure \ref{fig_contours_HR3}.




\subsection{$Mock\ galaxies\ for\ systematic\ correction$}

In the construction of HR4 mock galaxies \citep{hong2016}, 
the J08 model \citep{jiang2008}, a model derived from cosmological simulations, 
was adopted as the default option to calculate the merger timescale and to determine when a satellite galaxy is completely disrupted.

\cite{hong2016} also considered four alternative models for the calculation of merger timescale:
1) the LC93 model \citep{LC93} based on analytic calculation, 
2) the B08 \citep{B08} and 3) the V13 \citep{V13} models based on isolated simulations,
and 4) the M12 model \citep{M12} based on cosmological simulation.
Among the five merger timescale models, 
J08 model produces mock galaxies having properties and 2pCF best agree with the observational data \citep{hong2016};
the LC93 model, having the shortest merger timescale among the five models,
produces galaxies with properties and 2pCF most deviated from the observations.

\begin{figure*}
   \centering{
   \includegraphics[width=16cm]{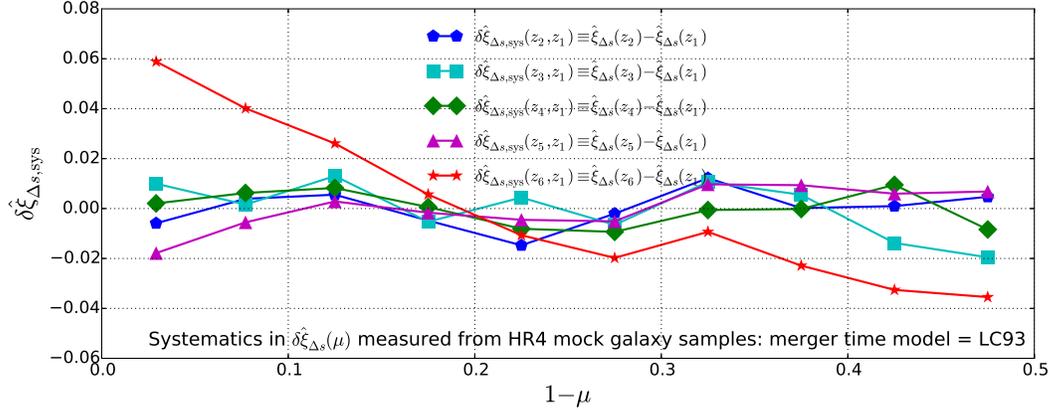}
   }
   \caption{\label{fig_sys_LC93}
   Systematics in $\delta \hat \xi_{\Delta s}$, measured from the HR4 mock galaxies
   with the LC93 model adopted as the merger timescale model.
   Compared with the main result using the merger timescale model J08 (Figure \ref{fig_sys}), 
   there is a $\sim 25\%$ change in the amplitude of $\delta \hat \xi_{\Delta s}$.
   }
\end{figure*}

We adopt mock surveys created using the four alternative models to
measure the systematic correction $\delta \hat\xi_{\Delta s, {\rm sys}}$
and derive the cosmological constraints.
The difference between the $\delta \hat\xi_{\Delta s, {\rm sys}}$ estimated from the J08 model and the other models is $5\%-25\%$.
As an example, Figure \ref{fig_sys_LC93} presents the systematics correction estimated using LC93,
which has a $\sim25\%$ difference from the J08 estimation.

Nevertheless, Figure \ref{fig_contour_RB} shows that, 
when the alternative models adopted, 
the cosmological constraints are consistent with our main results;
the discrepancy is $\lesssim0.3\sigma$.



\subsection{$Different\ s_{\rm max}$}

The results for $s_{\rm max}=30,\ 50 h^{-1}{\rm Mpc}$ are displayed in the first and second panels of the third row of Figure \ref{fig_contour_RB}.
Both of them are similar to the main results.


\subsection{$Cut\ on\ \mu$}

In our default options, we take $1 - \mu\ge0.01$.
For comparison, Figure \ref{fig_contour_RB} presents the cosmological constraints using the limits 
0.02, 0.05, and 0.1.
A larger limit means we abandon more angular regions near the LOS.
We find that changing the cut has little effect on the cosmological constraints.

This test suggests that our result is not affected by the fiber collisions and redshift failures.
If fiber collisions and redshift failures have any sizable effect,
since they mainly affect the angular region close to LOS,
one expects a systematic variation of the cosmological constraints when varying the limit on $1-\mu$.

\subsection{$Excluding\ the\ 6th\ redshift\ bin$}

The highest redshift data requires the largest systematic correction, 
so it would be interesting to know how much of an effect there is when it is removed from the analysis.

The results when excluding the 6th redshift bin from the analysis is shown in the last panel of Figure \ref{fig_contour_RB}.
The constraint becomes weaker.
The 1$\sigma$ contour is shifted towards large values of $\Omega_m$ and $w$ by $\lesssim0.4\sigma$,
which should not significantly affect the combined constraint.


\subsection{$Different\ s_{\rm min}$}


Among the different options, 
the one most likely to affect our results is $s_{\rm min}$.
The 2pCF has relatively large values on smaller scales
(e.g. for the CMASS galaxies $\xi\approx10,5,1,0.25,0.05$ at $s=2,4,10,20,40 h^{-1}{\rm Mpc}$).
Varying $s_{\rm min}$ can dramatically change the value of $\xi_{\Delta s}(\mu)$,
and thus is likely to have a large effect on the result.

A series of input-output tests are conducted to check whether the derived constraints depend on $s_{\rm min}$.
The linear regression of the $\delta\hat\xi_{\Delta s}$ (of the observational sample) 
measured in the WMAP5 cosmology ($\Omega_m=0.26$ $\Lambda$CDM) 
is adopted as $\delta\xi_{\Delta s, \rm sys}$. 
We then follow the procedures of Sec. \ref{sec:methodology} to derive cosmological constraint.

\begin{figure*}
   \centering{
   \includegraphics[width=5.7cm]{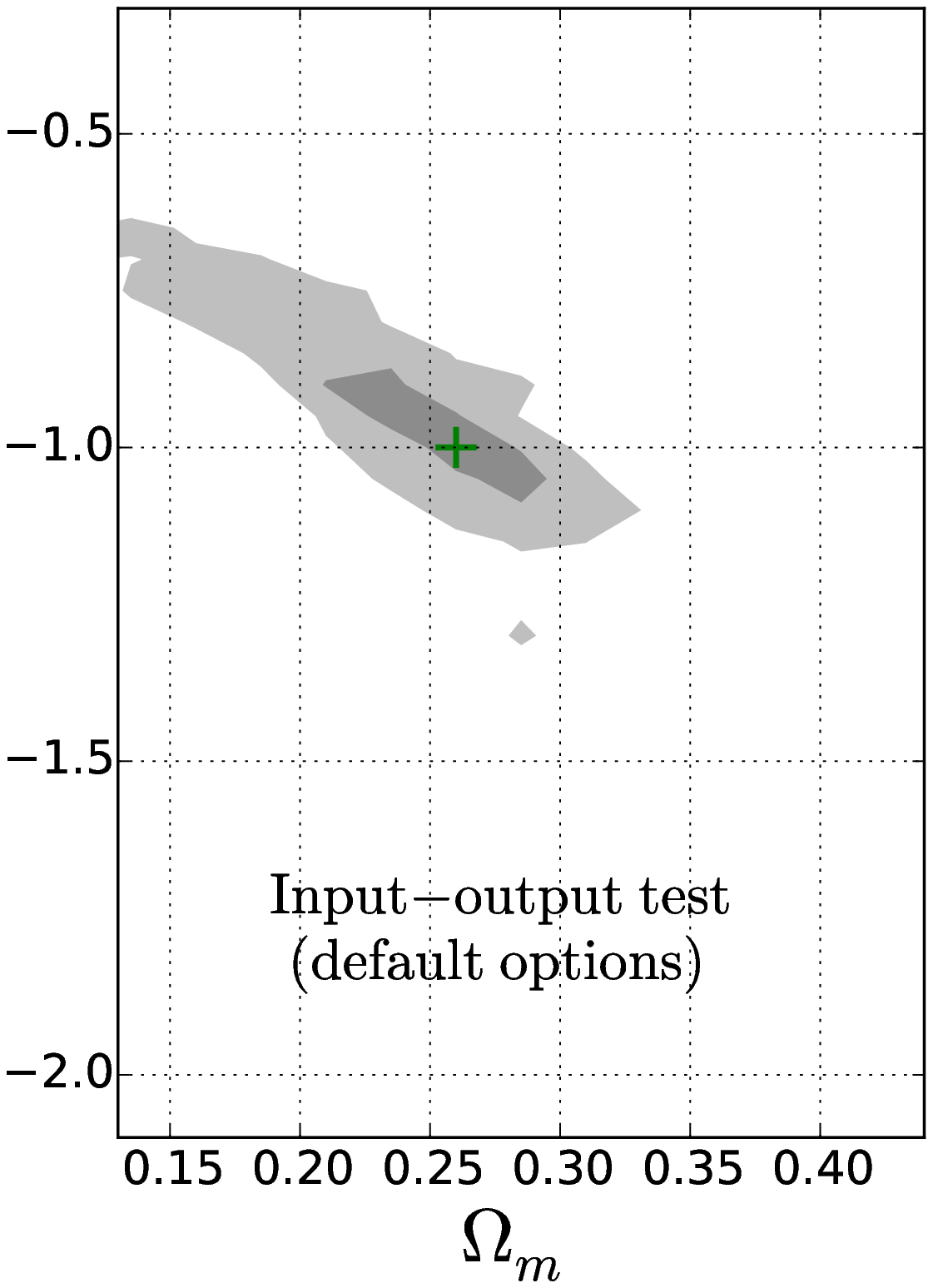}
   \includegraphics[width=10cm]{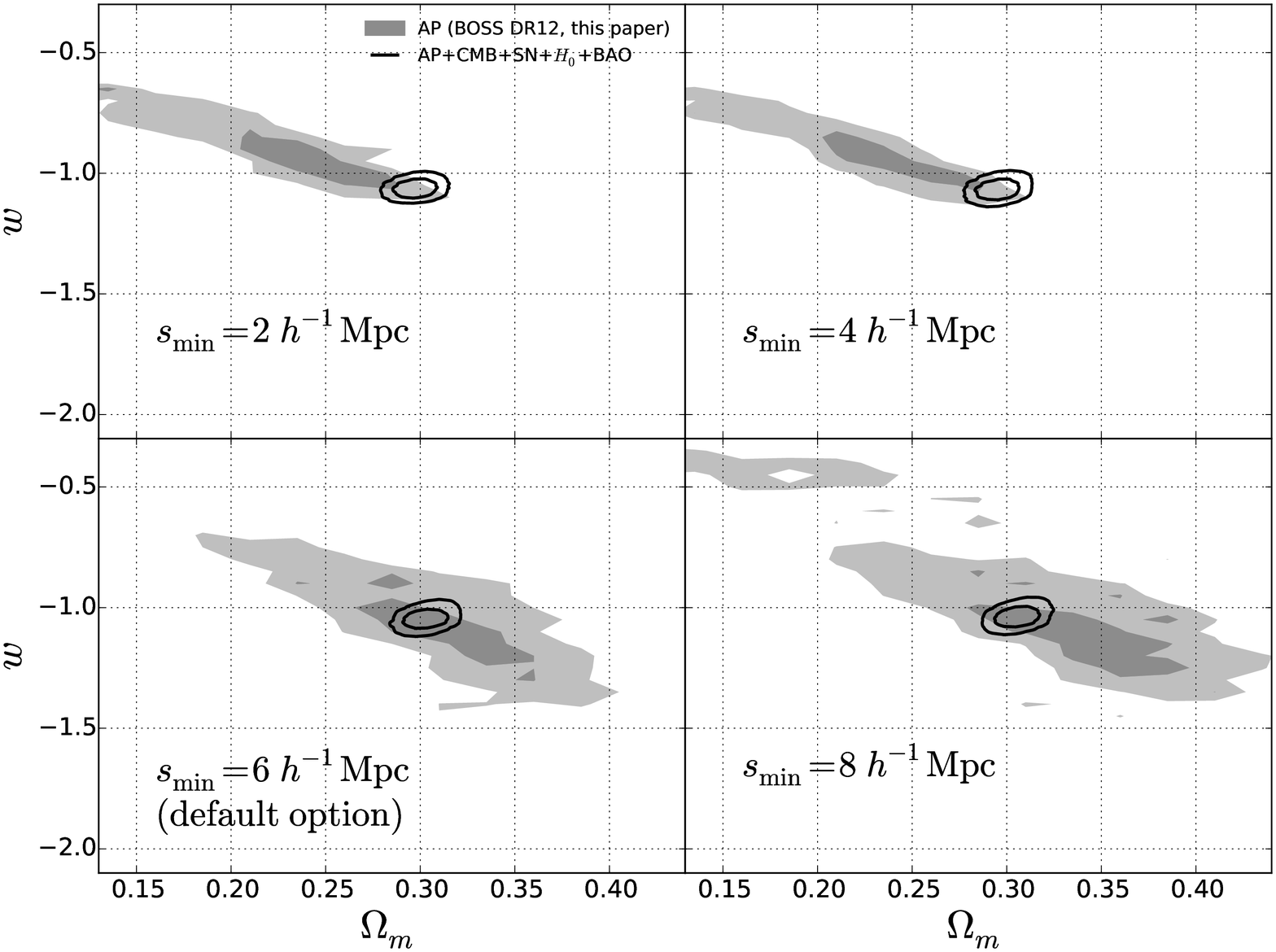}}
   \caption{ \label{fig_contour_diffsmin} 
   Left panel: The input-output test in case of using our default options.
   The derived cosmological constraints are fully consistent with the input WMAP5 cosmology (marked by green plus).
   Right panels: Likelihood contours (68\%, 95.4\%) in the $\Omega_m$-$w$ plane for using $s_{\rm min}=2,4,6,8\ h^{-1} {\rm Mpc}$.
   Gray filled contours are cosmological constraints derived using our AP analysis.
   For $s_{\rm min}=2,4h^{-1}{\rm Mpc}$, the contours are shifted to the upper-left corner, 
   but the deviation from the $s_{\rm min}=6h^{-1}{\rm Mpc}$ is $\lesssim1\sigma$.
   The black solid contour lines show the constraint when combined with CMB+BAO+JLA+$H_0$.
   The combined result is rather insensitive to choices of $s_{\rm min}$.}
\end{figure*}

When adopting $s_{\rm min}=6\ h^{-1}{\rm Mpc}$, the best-fit cosmology is exactly what we expect: the WMAP5 cosmology, 
i.e. the input cosmology is successfully recovered (left panel of Figure \ref{fig_contour_diffsmin}).
So $s_{\rm min}=6\ h^{-1}{\rm Mpc}$ is adopted as the default option in our analysis.
Other values of $s_{\rm min}$ do not recover the input cosmology as well as $s_{\rm min}=6h^{-1}{\rm Mpc}$.


The cosmological constraints obtained using $s_{\rm min}=2,4,6,8\ h^{-1} {\rm Mpc}$ 
are presented in the right panel of Figure \ref{fig_contour_diffsmin}.
In the case of $s_{\rm min}\neq6h^{-1} {\rm Mpc}$, we apply a simple correction to the derived likelihood map, 
based on the discrepancy between the best-fit cosmology and the input cosmology found in the input-output test.

These contours are consistent with each other within 1$\sigma$.
The $s_{\rm min}=8\ h^{-1}{\rm Mpc}$ result is very close to the main result.
The $s_{\rm min}=2$ or $4\ h^{-1}{\rm Mpc}$ contours show a systematical shift to the upper-left corner,
but the deviation is $\lesssim1\sigma$.
The difference choices of $s_{\rm min}$ produce little effect on 
the joint constraint when combined with other cosmological probes (CMB+BAO+JLA+$H_0$). 
The difference from the main results is $\lesssim0.3\sigma$.

The small scales contribute more in the integration of $\xi$ over $s$,
so the result is more affected by $s_{\rm min}$ compared with $s_{\rm max}$. 
We are going  to investigate the redshift evolution of 2d 2pCF in a future study, 
which hopefully can resolve this issue.


\subsection{$Summary$}


We conduct a series of tests and find that the derived cosmological constraints are insensitive to the options adopted in the analysis.
Cosmological constraints are derived using different options of $s_{\rm min}$, $s_{\rm max}$, limits on $1-\mu$, mocks for covariance estimation,
and satellite galaxy merger time scale.
We do not detect a significant change of our results.



Yet there are still unchecked options. 
One example is the redshift binning scheme.
In this analysis, the galaxies are split into six redshift bins of
$0.150<z_1<0.274<z_2<0.351<z_3<0.430<z_4<0.511<z_5<0.572<z_6<0.693$.
We have not checked whether the results depend on choices such as the redshift limits and the number of redshift bins,
because each redshift binning scheme requires to repeat the whole analysis that involves the calculation of the correlation function 
at each point of the parameter space.
This issue is worthy of investigation in future studies, 
to design a optimized scheme having the maximal ability to constrain cosmological parameters.

In summary, the tests show that, the cosmological constraints reported in this paper are robust.


\

\

\

\end{document}